\documentclass[twocolumn,nofootinbib,preprintnumbers,floatfix,aps,prd]{revtex4}

\usepackage[hyperfootnotes=false,bookmarks=false]{hyperref}
\usepackage{graphicx}
\usepackage{amsmath}
\usepackage{xcolor}

\newcommand{\eq}[1]{Eq.~\eqref{eq:#1}}
\newcommand{\eqs}[2]{Eqs.~\eqref{eq:#1} and \eqref{eq:#2}}

\renewcommand{\sec}[1]{Sec.~\ref{sec:#1}}

\newcommand{\subsec}[1]{Sec.~\ref{subsec:#1}}
\newcommand{\subsecs}[2]{Secs.~\ref{subsec:#1} and \ref{subsec:#2}}
\newcommand{\fig}[1]{Fig.~\ref{fig:#1}}
\newcommand{\figs}[2]{Figs.~\ref{fig:#1} and \ref{fig:#2}}
\newcommand{\app}[1]{App.~\ref{app:#1}}

\newcommand{\mycites}[1]{Refs.~\cite{#1}}
\newcommand{\mycite}[1]{Ref.~\cite{#1}}

\newcommand{\abs}[1]{\lvert#1\rvert}
\newcommand{\Abs}[1]{\bigl\lvert#1\bigr\rvert}
\newcommand{\ord}[1]{{\mathcal O}(#1)}

\newcommand{\ORD}[1]{{\mathcal O}\biggl(#1\biggr)}

\newcommand{\bare}{\mathrm{bare}}

\newcommand{\df}{\mathrm{d}}
\newcommand{\img}{\mathrm{i}}

\newcommand{\Tau}{\mathcal{T}}
\newcommand{\eps}{\epsilon}

\newcommand{\GeV}{\,\mathrm{GeV}}
\newcommand{\TeV}{\,\mathrm{TeV}}

\newcommand{\nn}{\nonumber}

\newcommand{\cL}{{\mathcal L}}
\newcommand{\cM}{{\mathcal M}}
\newcommand{\cI}{{\mathcal I}}

\newcommand{\Ecm}{E_\mathrm{cm}}
\newcommand{\lqcd}{\Lambda_\mathrm{QCD}}
\newcommand{\cusp}{\mathrm{cusp}}
\newcommand{\cut}{\mathrm{cut}}
\newcommand{\jet}{\mathrm{jet}}
\newcommand{\FO}{\mathrm{FO}}
\newcommand{\cm}{\mathrm{cm}}
\newcommand{\nons}{\mathrm{nons}}
\newcommand{\resum}{\mathrm{resum}}
\newcommand{\sing}{\mathrm{sing}}
\newcommand{\run}{\mathrm{run}}
\newcommand{\central}{\mathrm{central}}
\newcommand{\vary}{\mathrm{vary}}

\newcommand{\TauB}{\Tau_B}
\newcommand{\TauC}{\Tau_C}
\newcommand{\TauBcm}{\Tau_{B\mathrm{cm}}}
\newcommand{\TauCcm}{\Tau_{C\mathrm{cm}}}

\newcommand{\msb}{{\overline{\rm MS}}}

\allowdisplaybreaks[2]

\begin{document}


\preprint{\vbox{
\hbox{DESY 14-240}
}}

\title{Rapidity-Dependent Jet Vetoes}

\author{Shireen Gangal}
\author{Maximilian Stahlhofen}
\author{Frank J.~Tackmann}
\affiliation{Theory Group, Deutsches Elektronen-Synchrotron (DESY), D-22607 Hamburg, Germany\vspace{0.5ex}}

\date{December 15, 2014}

\begin{abstract}

Jet vetoes are a prominent part of the signal selection in various analyses at the LHC.
We discuss jet vetoes for which the transverse momentum of a jet is
weighted by a smooth function of the jet rapidity.
With a suitable choice of the rapidity-weighting function, such jet-veto variables can be factorized and
resummed allowing for precise theory predictions.
They thus provide a complementary way to divide phase space into
exclusive jet bins. In particular, they provide a natural and theoretically clean
way to implement a tight veto on central jets with the veto constraint getting looser
for jets at increasingly forward rapidities.
We mainly focus our discussion on the $0$-jet case in color-singlet processes, using
Higgs production through gluon fusion as a concrete example.
For one of our jet-veto variables we compare the resummed theory prediction at NLL$'+$NLO with
the recent differential cross section measurement by the ATLAS experiment in the $H\to\gamma\gamma$ channel, finding good agreement.
We also propose that these jet-veto variables can be measured and tested against theory predictions in other SM processes,
such as Drell-Yan, diphoton, and weak diboson production.

\end{abstract}

\maketitle


\section{Introduction}
\label{sec:intro}

Jet vetoes play an important role at the LHC in Higgs property measurements as well as in searches for physics beyond the Standard Model. They are utilized to reduce backgrounds and more generally are used to classify the data into exclusive categories, ``jet bins'', based on the number of hadronic jets in the final state. The default jet variable by which jets are currently classified and vetoed is the transverse momentum $p_{Tj}$ of a jet.

While a veto on additional jets can be desirable in many contexts, the application of a tight jet veto is usually subject to both theoretical and experimental limitations. Theoretically, applying a tight jet veto leads to Sudakov double logarithms of the jet-veto variable in perturbation theory, which as the veto gets tighter (smaller veto cuts) become larger and dominate the perturbative series, leading to increased theoretical uncertainties in the fixed-order (FO) predictions~\cite{Stewart:2011cf}. This can be remedied by systematically resumming the jet-veto logarithms to all orders~\cite{Stewart:2009yx, Stewart:2010tn, Berger:2010xi, Banfi:2012yh, Becher:2012qa, Tackmann:2012bt, Banfi:2012jm, Liu:2012sz, Liu:2013hba, Becher:2013xia, Stewart:2013faa, Boughezal:2013oha, Shao:2013uba, Li:2014ria, Boughezal:2014qsa, Jaiswal:2014yba}, provided that the considered jet-veto variable is resummable and under good enough theoretical control.

Experimentally, jets can only be robustly reconstructed down to some minimum $p_T$, which limits how low one can go in the jet veto cut, i.e., how tight one can make the jet veto. Furthermore, in harsh pile-up conditions low-$p_T$ jets are particularly hard to identify at forward rapidities (beyond $\abs{\eta} \gtrsim 2.5$), when a large part or all of the jet area lies in a detector region where no tracking information is available.

In principle, one possibility would be to place a hard cut on the (pseudo)rapidity $\eta_j$ of the classified jets, i.e., one only considers and possibly vetoes jets within a certain range of central rapidities, $\abs{\eta_j} < \eta^\cut$. Theoretically, such a hard rapidity cut represents a nonglobal measurement and changes the logarithmic structure~\cite{Tackmann:2012bt}. This means that a priori it is not clear how to consistently incorporate it into the jet-veto resummation at higher orders, and none of the present jet-veto resummations for $p_{Tj}$ actually includes such a rapidity cut. (In Monte Carlo studies, a cut at $\eta^\cut\sim 2.5$ has an $\ord{10\%}$ effect on the cross section for typical $p_{Tj}$ vetoes~\cite{Berger:2010xi, Banfi:2012yh}.) Another option, which avoids a hard rapidity cut, is to raise the cut on $p_{Tj}$, and thus loosen the jet veto everywhere. Clearly, this may also not be ideal since one now looses the utility of a tight jet veto for central jets.

In this paper, we discuss a class of jet-veto variables which explicitly depend on the jet rapidity $y_j$,
\begin{equation}
\Tau_{fj} = p_{Tj} \,f(y_j)
\,,\end{equation}
where $f(y_j)$ is some weighting function of $y_j$.
(The difference between $\eta_j$ and $y_j$ due to a nonzero jet mass is not relevant for now and either could be used. We will come back to this at the end of \sec{variables}.)

By classifying jets according to $\Tau_{fj}$ and only allowing jets with $\Tau_{fj} < \Tau^\cut$, we effectively have a rapidity-dependent veto on $p_{Tj}$,
\begin{equation}
p_{Tj} < \frac{\Tau^\cut}{f(y_j)}
\,.\end{equation}
If the weighting function $f(y)$ is chosen as a decreasing function of $\abs{y}$ this corresponds to a veto which gets tighter at central rapidities and looser at forward rapidities. Effectively, the contribution of forward jets is then smoothly suppressed by the weighting function $f(y_j)$. At the same time, $f(y_j)$ can be chosen such that explicit theoretical control is maintained. In fact, all the variables we discuss can be resummed to a similar (and possibly higher) level of precision as $p_{Tj}$. In this way, one can largely avoid the theoretical and experimental limitations discussed above. (Of course, the lowest $\Tau_{fj}$ values that can be measured are ultimately still limited by how well central jets can be measured.)

Apart from such practical considerations, given the usefulness of jet binning, it is clearly beneficial to have several alternative and complementary ways to perform it, as this gives the experiments a wider range of options for optimizing their analyses. One could even optimize the form of $f(y)$ to the needs of a given analysis. On the theoretical side, it allows one to test jet-veto resummations in different and as of now unexplored regimes.

Note that two special cases we have already discussed above are no weighting, $f(y) \equiv 1$, for which $\Tau_{fj} \equiv p_{Tj}$, while $f(y) = \theta(\abs{y} < y^\cut)$ is equivalent to a hard cut on the jet rapidity.

In the next section, we discuss in more detail possible weighting functions and jet-veto variables. We introduce two types of weighting functions, corresponding to the jet-veto variables $\TauB^\jet$ and $\TauC^\jet$ (and variants of them defined in different kinematic frames), whose factorization and resummation are discussed in detail in \sec{theory}. (The $\TauB^\jet$ variable has been discussed before in \mycite{Tackmann:2012bt}.) In \sec{results}, we then provide numerical predictions at NLL$'+$NLO order for gluon-fusion Higgs$+$0-jet cross sections with $\Tau_B^\jet$ and $\Tau_C^\jet$-type jet vetoes. The differential cross section in bins of $\TauC^\jet$ was measured recently by the ATLAS $H\to\gamma\gamma$ analysis~\cite{Aad:2014lwa}, and we compare our predictions with the experimental measurements, finding good agreement. We conclude in \sec{conclusions}.

\section{Jet-Veto Variables}
\label{sec:variables}

\begin{figure*}[t!]
\begin{center}
\includegraphics[width=\columnwidth]{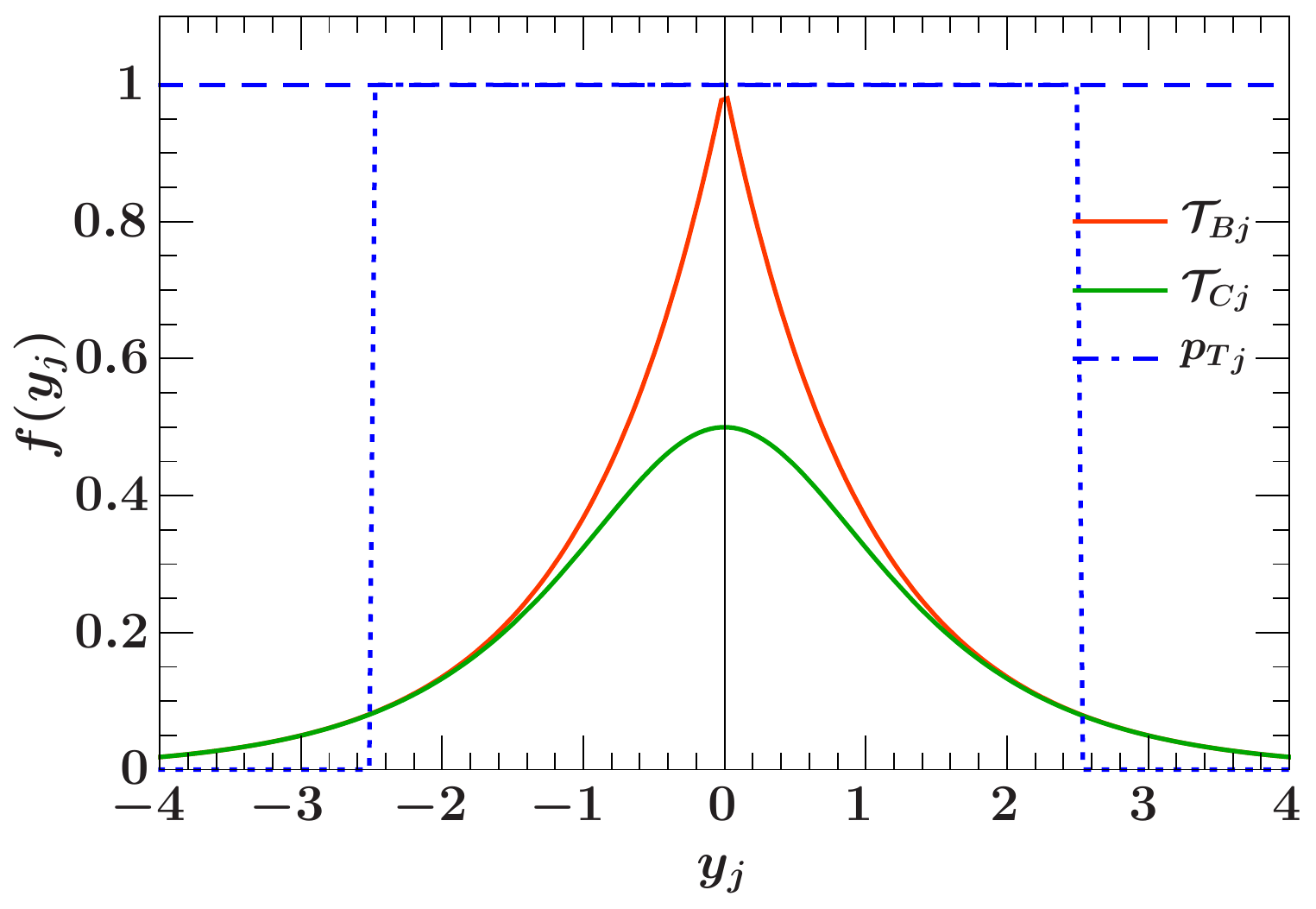}%
\hfill%
\includegraphics[width=\columnwidth]{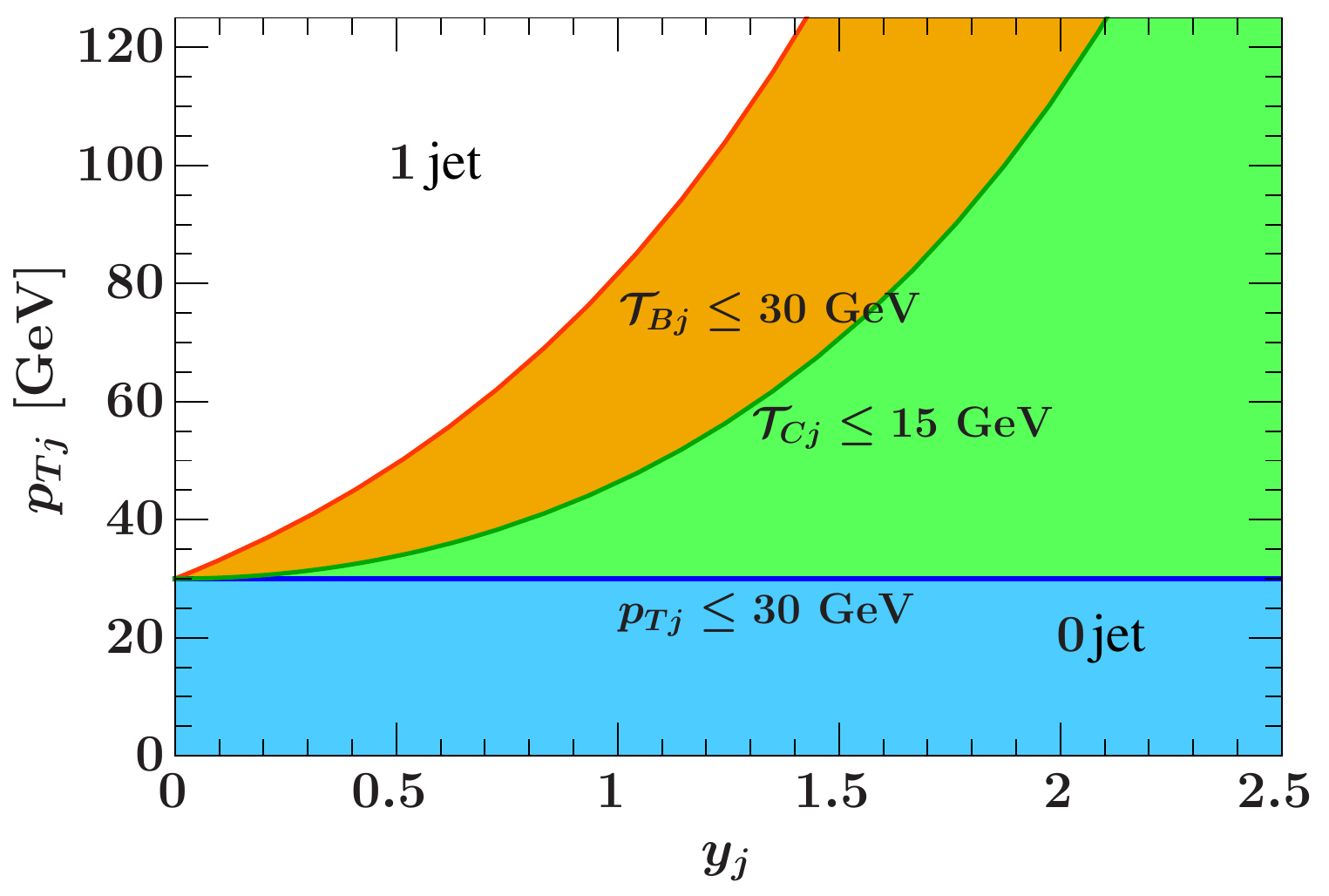}%
\end{center}
\vspace{-1ex}
\caption{Left: Illustration of rapidity weighting functions for $\Tau_{Bj}$ (orange), $\Tau_{Cj}$ (green), and $p_{Tj}$ (blue dashed). The blue dotted lines show a fixed cut on the jet rapidity. Right: Phase-space region in the $p_{Tj}-y_j$ plane selected by the different jet-veto variables. (Here we take $Y=0$, so $\Tau_{Bj}=\Tau_{B\cm j}$.)}
\label{fig:phasespace}
\end{figure*}

In general, one can distinguish two classes of variables, inclusive and jet-based, that can be used to classify the number of jets; see \mycite{Tackmann:2012bt} for a detailed discussion. Inclusive variables, such as beam thrust or $N$-jettiness~\cite{Stewart:2009yx, Stewart:2010tn}, do not depend on a specific jet algorithm or jet size. Instead, they sum over all hadrons in the final state, and provide a global view of the event, effectively measuring the sum of all emissions.
On the other hand, jet-based (exclusive) variables are based on identifying jets $J(R)$ with radius $R$ using a specific jet algorithm. They provide a local view of the event and measure emissions locally with an effective ``resolution'' size set by $R$.

In this paper, we focus on jet-based variables, since they are more straightforward to use experimentally. However, on the theory side, the jet-algorithm dependence renders their resummation structure more involved at higher orders. We will comment on this later on in \subsecs{fact}{resum}.

Given the set of jets, $J(R)$, identified by some jet clustering algorithm, we define
\begin{equation} \label{eq:pTjet}
p_T^\jet = \max_{j\in J(R)}\, p_{Tj}
\end{equation}
as the largest $p_{Tj}$ of any jet. Requiring $p_T^\jet < p_T^\cut$ vetoes any event having at least one jet with $p_{Tj} > p_T^\cut$. The so-defined $0$-jet cross section then consists of events where all jets have $p_{Tj} < p_T^\cut$. It is important to note that despite that fact, this does not actually require one to reconstruct jets with $p_{Tj} < p_T^\cut$. Rather, one only has to be able to reconstruct jets with $p_{Tj} > p_T^\cut$ which are to be vetoed. The resummation for a veto on $p_T^\jet$ is known to NNLL and partially beyond~\cite{Banfi:2012yh, Becher:2012qa, Tackmann:2012bt, Banfi:2012jm, Becher:2013xia, Stewart:2013faa}.

For simplicity, we explicitly consider the $0$-jet bin in the following. The extension to an $N$-jet bin is obtained by simply removing from the set $J(R)$ the $N$ identified jets that have been selected as the ``signal'' jets. (The signal jets do not necessarily have to be selected as the $N$ jets with the highest $p_{Tj}$, but one can use whatever kinematic selection and/or flavor-tagging is appropriate for the hard signal process of interest.) Doing so then defines $p_T^\jet$ as the largest $p_{Tj}$ of any additional unwanted jet (i.e. from additional initial-state or final-state radiation), which are to be vetoed by requiring $p_T^\jet < p_T^\cut$.

We can generalize this to $\Tau_{fj}$ by defining
\begin{align} \label{eq:exclvarTau}
\Tau_f^\jet &= \max_{j\in J(R)}\, \Tau_{fj} = \max_{j\in J(R)}\, \abs{\vec{p}_{Tj}}\, f(y_j)
\,.\end{align}
We now distinguish between the $\Tau_{fj}$ value of any given jet $j$ and $\Tau_f^\jet$, which is the maximum $\Tau_{fj}$ of all jets (or all additional jets for the case of $N$ selected signal jets). In particular, the ``leading'' jet is now determined by $\Tau_{fj}$ and not by $p_{Tj}$.%
\footnote{In principle, one could also measure the $\Tau_{fj}$ of the leading-$p_T$ jet. However, using this as a jet veto would make things much more involved and we will not consider such mixed cases.}
We can then classify events into jet bins according to $\Tau_f^\jet$ and define a $0$-jet cross section by requiring
\begin{equation}
\Tau_f^\jet < \Tau^\cut
\,,\end{equation}
which consists of events where all jets have $\Tau_{fj} < \Tau^\cut$. The corresponding inclusive $1$-jet cross section defined by requiring $\Tau_f^\jet > \Tau^\cut$ consists of all remaining events that have at least one jet with $\Tau_{fj} > \Tau^\cut$. Similar to the $p_T^\jet$ case, this $\Tau_f^\jet$ binning now requires one to be able to reconstruct jets down to $\Tau_{fj} > \Tau^\cut$, while jets below $\Tau^\cut$ do not have to be reconstructed.

The four jet-veto variables we consider in the following  are defined with their respective weighting functions as follows:
\begin{align}
\TauB: \qquad f(y) &= e^{-\abs{y - Y}}
\,, \label{eq:TauB}\\ \qquad
\TauBcm: \qquad f(y) &= e^{-\abs{y}}
\,, \label{eq:TauBcm} \\ \qquad
\TauC: \qquad f(y) &= \frac{1}{2\cosh(y - Y)}
\,, \label{eq:TauC} \\ \qquad
\TauCcm: \qquad f(y) &= \frac{1}{2\cosh y}
\label{eq:TauCcm}
\end{align}
Here, $Y$ denotes the rapidity of the hard system. For the $0$-jet case, this is equivalent to the nonhadronic final state, i.e. $Y$ is the vector-boson rapidity for Drell-Yan or the Higgs-boson rapidity in gluon-fusion Higgs production. By including $Y$ in $\TauB$ and $\TauC$, the variables become longitudinally boost-invariant.%
\footnote{They can be thought of as being defined in the frame where $Y = 0$, and in all other frames by boosting from that frame.}
\pagebreak[4]
On the other hand, $\TauBcm$ and $\TauCcm$ are explicitly defined in the hadronic center-of-mass (cm) frame, i.e. the lab frame, which has the advantage that one does not have to reconstruct $Y$.

The different rapidity weighting functions are illustrated in the left panel of \fig{phasespace} by the orange ($\Tau_{Bj}$) and green ($\Tau_{Cj}$) lines. For comparison, the blue dashed line shows the case of $p_{Tj}$ ($f(y) = 1$) and the blue dotted line a hard rapidity cut. The weighting $\sim e^{-\abs{y}}$ for $\Tau_{Bj}$ is the same as that for inclusive beam thrust, so we can think of $\Tau_{Bj}$ as the beam thrust of a single jet and $\TauB^\jet$ as the maximum jet beam thrust (which was first discussed in Ref.~\cite{Tackmann:2012bt}). The rapidity weighting for $\Tau_{Cj}$ is the same as that for the $C$-parameter event-shape in $e^+e^-\to$ dijets. It becomes equal to $\Tau_{Bj}$ at forward rapidities, while at central rapidities it is much flatter and approaches $p_{Tj}/2$ for $y_j = 0$. Experimentally, this has the advantage that $\TauC^\jet$ can be measured to much smaller values. The region in the $p_{Tj}-y_j$ phase space selected by the different variables is illustrated in \fig{phasespace} on the right. The lines correspond to the given fixed value of $\Tau_{fj}$. They separate the ``0-jet'' region (colored), where the jet would be allowed by the corresponding jet-veto cut, and the ``1-jet'' region (uncolored), where the jet would be vetoed.

The strict exponential weighting for $\Tau_{Bj}$ is distinguished by the fact that $\Tau_{Bj}$ is related to the small light-cone component with respect to the beam axis of the total jet momentum. More precisely, including the nonzero mass of the jet, $m_j$, we have
\begin{align}
p_{Tj}\, e^{-\abs{\eta_j}} &= \abs{\vec p_{j}}  - \abs{p_{zj}}
\nn \,, \\
m_{Tj}\, e^{-\abs{y_j}}
&\equiv \sqrt{p_{Tj}^2 + m_j^2}\, e^{-\abs{y_j}} = E_j - \abs{p_{zj}}
\,.\end{align}
Either of these variants can be used as alternative definitions of $\Tau_{B\cm}$ (and analogously for $\Tau_{B}$), if this is desired or turns out to be advantageous for their experimental measurement. Theoretically, all of these are distinct variables which however have a very similar logarithmic structure at small $\TauB^\jet$. The different treatment of the jet mass amounts to having different jet clustering corrections for each variable and can be taken into account systematically. They start entering at $\ord{\alpha_s^2}$, which is beyond the order we will work at for our numerical results in \sec{results}.

The analogous discussion holds for $\Tau_{C(\cm)}$, which including nonzero $m_j$ can be defined in terms of either combination of $p_{Tj}$ or $m_{Tj}$ and $\eta_j$ or $y_j$. Explicitly,
\begin{align}
\frac{p_{Tj}}{2\cosh \eta_j} &= \frac{p_{Tj}^2}{2 \abs{\vec p_j}}
\nn \,, \\
\frac{m_{Tj}}{2\cosh y_j} &= \frac{p_{Tj}^2 + m_j^2}{2 E_j}
\,.\end{align}
Again, either of these variants could be used as alternative definitions of $\Tau_{Cj}$. The ATLAS measurement~\cite{Aad:2014lwa} uses the last variant above in the $Y=0$ frame, i.e., $\Tau_{Cj}\equiv m_{Tj}/[2\cosh(y_j - Y)]$.

Note that $1/(2\cosh x) = 1/(e^x + e^{-x}) \to e^{-\abs{x}}$ for large $\abs{x}$, such that at forward rapidities $\Tau_{Cj}$ has the same behavior as $\Tau_{Bj}$, as seen in \fig{phasespace}. For this reason, its logarithmic structure is closely related to that of $\Tau_{Bj}$, and in particular the same technology can be used to resum it to the same level of accuracy.%
\footnote{Analogously, in the context of $e^+e^-\to$ jets, the $C$-parameter event shape is closely related to thrust, which makes it comparably easy to resum to the same high order as for thrust~\cite{Alioli:2012fc, Hoang:2014wka}.}
The same reasoning also applies more generally to any (continuous) weighting function $f(y)$ that approaches $e^{-\abs{y}}$ at large rapidities. This gives considerable freedom in choosing other alternative rapidity weighting functions yielding resummable jet-veto variables.

\section{Theory predictions for Higgs production}
\label{sec:theory}

In this Section, we discuss the theory predictions for the rapidity-weighted jet vetoes. We first discuss their general factorization and resummation using soft-collinear effective theory (SCET)~\cite{Bauer:2000ew, Bauer:2000yr, Bauer:2001ct, Bauer:2001yt, Bauer:2002nz, Beneke:2002ph}. This discussion applies to any color-singlet process. However, to be concrete, we phrase it in the context of $gg\to H$ production, for which we also obtain explicit numerical results at NLL$'+$NLO.

\subsection{Factorization}
\label{subsec:fact}

The full $H+0$-jet cross section differential in the Higgs rapidity $Y$ and with a cut on $\Tau_f^\jet < \Tau^\cut$ can be written as
\begin{align} \label{eq:fullXsec}
\frac{\df\sigma_0}{\df Y}(\Tau_f^\jet \!< \Tau^\cut)
&= \frac{\df\sigma_0^\resum}{\df Y}(\Tau_f^\jet \!< \Tau^\cut)
\nn \\ & \quad
+ \frac{\df\sigma_0^\nons}{\df Y}(\Tau_f^\jet \!< \Tau^\cut)
\,,\end{align}
where the first term contains the resummed logarithmic contributions, which dominate at small $\Tau^\cut$, and the second term represents the ``nonsingular'' corrections, which are suppressed relative to the leading terms by $\ord{\Tau^\cut/m_H}$ and vanish in the limit $\Tau^\cut\to 0$. The cross sections also depends on the jet algorithm and jet radius $R$, which we suppress here to keep the notation simple.

The all-order factorization for $\TauB^\jet$ was discussed in detail in \mycite{Tackmann:2012bt}. It relies on the fact that the measurement function, $\cM_f^\jet$, for a veto on a jet-based observable $\Tau_f^\jet<\Tau^\cut$ can be expressed as a simple product of measurement functions on the individual jets,
\begin{align} \label{eq:Mjet}
\cM^\jet_f(\Tau^\cut)
= \theta(\Tau_f^\jet < \Tau^\cut)
= \prod_{j\in J(R)} \theta(\Tau_{fj}<\Tau^\cut)
\,.\end{align}
This holds for any $\Tau_f^\jet$, and in particular for the four variables defined in Eqs.~\eqref{eq:TauB}-\eqref{eq:TauCcm}.
We can furthermore explicitly disentangle the measurements acting on the different collinear and soft sectors in the effective theory by writing
\begin{align} \label{eq:Mfac}
\cM^\jet_f(\Tau^\cut) &= \cM_{fa}^\jet(\Tau^\cut)\, \cM_{fb}^\jet(\Tau^\cut)\, \cM_{fs}^\jet(\Tau^\cut)
\nn \\ &\quad
+ \delta\cM_f^\jet(\Tau^\cut)
\,,\end{align}
where the $\cM^\jet_{fi}$ for $i=a,b,s$ are defined as the measurement $\cM^\jet_f$ acting only on $n_a$-collinear, $n_b$-collinear ($n_a$ and $n_b$ being lightlike vectors along the two beam directions) and soft final state particles, respectively. Furthermore, $\delta\cM^\jet_f$ encodes the contributions to the full measurement where the jet algorithm clusters both soft and collinear emissions into the same jet, which inhibits the complete all-order factorization of the measurement function.
Since such contributions arise from independent emissions, they are suppressed by $\ord{R^2}$~\cite{Banfi:2012yh, Tackmann:2012bt}.

Hence, for $R^2\ll1$, the resummed contribution for the $\Tau_{B,C}^\jet < \Tau^\cut$ veto can be factorized to all orders in perturbation theory into hard ($H$), beam ($B$) and soft functions ($S$) as
\begin{align} \label{eq:TauBCfacto}
&\frac{\df\sigma_0^\resum}{\df Y}(\Tau_{B,C}^\jet \!< \Tau^\cut)
\nn \\ & \quad
= \sigma_B H_{gg}(m_t, m_H^2, \mu)\, B_g(m_H \Tau^{\cut},x_a,R,\mu)
\nn \\ &\quad\qquad \times
B_g(m_H \Tau^{\cut},x_b,R,\mu)\, S_{gg}^{B,C}(\Tau^{\cut},R,\mu)
\nn \\ &\qquad
+ \frac{\df\sigma_0^{\text{Rsub}}}{\df Y}(\Tau_{B,C}^\jet \!< \Tau^\cut, R)
\,,\end{align}
where
\begin{align}
x_{a,b} = \frac{m_H}{\Ecm}\,e^{\pm Y}
\,,\quad
\sigma_B = \frac{\sqrt{2} G_F\, m_H^2}{576 \pi \Ecm^2}
\,.\end{align}
The hard function is observable independent and is determined by the IR-finite part of the $\msb$ renormalized $ggH$ form factor, $C_{ggH}$, given in \eq{CggH},
\begin{align}
H_{gg}(m_t, m_H^2, \mu) = |C_{ggH}(m_t, m_H^2, \mu)|^2
\,.\end{align}
The only difference between $\TauB^\jet$ and $\TauC^\jet$ to all orders is their dependence on different soft functions, $S_{gg}^{B,C}$. The beam functions $B_i$ are the same for both observables, because they describe the effects of collinear initial-state radiation, i.e. emissions with forward rapidities, where the $\TauB^\jet$ and $\TauC^\jet$ measurements are equal up to power corrections (cf. left panel in \fig{phasespace}). This can be seen explicitly by expressing the variables $\Tau_{B,Cj}$ in terms of plus and minus momenta in one of the two collinear sectors,
\begin{equation}
 \Tau_{Bj} = p^+_j
 \,,\qquad
 \Tau_{Cj} = \frac{p^+_j p^-_j}{p^+_j + p^-_j}
 \,,\end{equation}
with $p^-_j \gg p^+_j$ and therefore $\Tau_{Cj}=\Tau_{Bj}+\ord{p^+_j/p^-_j}$, where $p^+_j/p^-_j \sim \Tau/m_H$ is a power correction.

The $\ord{R^2}$ corrections from $\delta\cM^\jet_f$ in \eq{Mfac} are collected in the $\df\sigma_0^{\text{Rsub}}$ piece in \eq{TauBCfacto}. They start contributing at $\ord{\alpha_s^2}$ and NNLL, and so are not yet needed at NLL$'$.
Note that another source of possibly factorization-violating contributions for hadronic observables is related to the interaction between spectator particles mediated by Glauber modes. Usually, these are not considered in perturbative predictions of jet cross sections. As argued in \mycite{Gaunt:2014ska}, for jet-based observables these effects are suppressed at least as $\ord{R^2}$, and we therefore also neglect them here.

For $\TauBcm^\jet$ and $\TauCcm^\jet$, the resummed contribution obeys a similar factorization of the form
\begin{align} \label{eq:TauBCcmfacto}
&\frac{\df\sigma_0^\resum}{\df Y}(\Tau_{B\cm,C\cm}^\jet \!< \Tau^\cut)
\nn \\ & \quad
= \sigma_B H_{gg}(m_t, m_H^2, \mu) \, B_g(m_H \Tau^{\cut} e^Y\!,x_a,R,\mu)
\nn \\ &\quad\qquad \times
B_g(m_H \Tau^{\cut} e^{-Y}\!,x_b,R,\mu)\, S_{gg}^{B,C}(\Tau^{\cut},R,\mu)
\nn \\ &\qquad
+ \frac{\df\sigma_0^{\text{Rsub}}}{\df Y}(\Tau_{B\cm,C\cm}^\jet \!< \Tau^\cut, R)
\,.\end{align}
Here, the hard, beam, and soft functions are the same as in \eq{TauBCfacto}, the only difference is that the beam functions are evaluated at different arguments. The soft function $S_{gg}^{B,C}$ is precisely the same as in \eq{TauBCfacto} because
it is a vacuum matrix element and, unlike the beam functions that involve the incoming proton states, has no reference to the frame other than through the measurement function itself. By a change (boost) of the soft integration momenta, the soft measurement function for $\Tau_{B\cm,C\cm}^\jet$ can therefore always be transformed into that for $\Tau_{B,C}^\jet$, respectively.
(Technically, this is a consequence of the RPI-III invariance~\cite{Chay:2002vy, Manohar:2002fd} of the soft function.)

The beam functions on the other hand are related to the ones in \eq{TauBCfacto} by a simple rescaling of $\Tau^{\cut}$. 
This is because for jets made of $n_a$($n_b$)-collinear particles we have $y_j>0$ ($y_j<0$) in the $Y = 0$ frame and correspondingly  $y_j>Y$ ($y_j<Y$) in the lab frame.
According to Eqs.~\eqref{eq:TauB}-\eqref{eq:TauCcm} for $n_{a,b}$-collinear jets we therefore have $\Tau_{C\cm}=\Tau_{B\cm}=\Tau_B e^{\mp Y}$, respectively. A detailed discussion of the analogous frame-dependence for inclusive beam thrust can be found in \mycite{Stewart:2009yx}.

\subsection{Resummation}
\label{subsec:resum}

The factorization in \eqs{TauBCfacto}{TauBCcmfacto} allows a consistent resummation of large logarithms $\sim \alpha_s^n \ln^m(\Tau^\cut/m_H)$ to all orders in $\alpha_s$.
To compute the corresponding resummation factors we have to derive and solve separate renormalization group equations (RGEs) in SCET for the hard, beam, and soft functions.

As a consequence of the simple multiplicative structure of the cross sections for the jet-based variables (as opposed to the convolutions for the respective jet-algorithm independent inclusive variables), also the RGEs take a product form,
\begin{align} \label{eq:anomdims}
\mu \frac{\df}{\df\mu} \ln\bigl[C_{ggH} (m_t,m_H^2,\mu)\bigr]
&= \gamma_{H}^g(m_H^2,\mu)
\,, \nn\\
\mu \frac{\df}{\df\mu} \ln\bigl[B_g (t^\cut,x,R,\mu) \bigr]
&= \gamma_{B}^g(t^\cut,R,\mu)
\,, \nn\\
\mu \frac{\df}{\df\mu} \ln\bigl[ S_g^{B,C} (\Tau^\cut\!,R,\mu) \bigr]
&= \gamma_{S}^g(\Tau^\cut,R,\mu)
\,,\end{align}
where in the beam function $t^{\cut}=m_H \Tau^{\cut}$ and $t^{\cut}=m_H \Tau^{\cut} e^{\pm Y}$ for Eqs.~\eqref{eq:TauBCfacto} and \eqref{eq:TauBCcmfacto}, respectively.

The generic all-order structure of the anomalous dimensions, as the sum of a noncusp part and an explicitly $\mu$-dependent cusp part, is fixed by RG invariance of the cross section and the well-known Sudakov form of the hard function, which is completely independent of the specific observable. Specifically, we have
\begin{align} \label{eq:anomdimsstructure}
\gamma_H^g(m_H^2,\mu) &= \Gamma_\cusp^g[\alpha_s(\mu)] \ln{\frac{-m_H^2\!-\!{\rm i}0}{\mu^2}} + \gamma_H^g[\alpha_s(\mu)]
\,, \nn \\
\gamma_B^g(t^\cut\!,R,\mu) &= -2 \Gamma_\cusp^g[\alpha_s(\mu)] \ln{\frac{t^\cut}{\mu^2}} + \gamma_B^g[\alpha_s(\mu), R]
\,,\nn \\
\gamma_S^g(\Tau^{\cut}\!,R,\mu) &= 4\Gamma_\cusp^g[\alpha_s(\mu)] \ln {\frac{\Tau^{\cut}}{\mu}} + \gamma_S^g[\alpha_s(\mu), R]
\,.\end{align}
The RG invariance of the cross section moreover requires that the soft anomalous dimension of $S_{gg}^{B}$ and $S_{gg}^{C}$ is the same to all orders in perturbation theory, and hence it is the same for all four observables we consider.

Note that beyond one loop (i.e. starting at NNLL and beyond the NLL$'$ order we will be interested in here) the beam and soft functions as well as their noncusp anomalous dimensions, $\gamma_{B,S}^g(\alpha_s, R)$, acquire an explicit $R$-dependence, as denoted in \eqs{anomdims}{anomdimsstructure} above. In particular, they receive corrections proportional to powers of $\ln R$~\cite{Tackmann:2012bt}, associated with the jet clustering. For too small $R$, these $\ln R$ clustering logarithms spoil the NLL precision, since they should formally be included in the logarithmic counting. (For $p_T^\jet$, the higher-order clustering logarithms have been discussed in Refs.~\cite{Alioli:2013hba, Dasgupta:2014yra} and for typical $R \gtrsim 0.4$ seem to be under sufficient control to not spoil the accuracy of the final predictions.)

Solving the above RGEs we get,
\begin{align} \label{eq:Bresum}
B_g(t^{\cut}\!,x,R,\mu) = U_B(t^{\cut}\!,\mu_B, \mu)\, B_g(t^{\cut}\!,x,R,\mu_B)
\end{align}
with the evolution factor given by
\begin{align} \label{eq:evolveB}
U_B(t^\cut\!,\mu_B,\mu) &= e^{K_B(\mu_B, \mu)}\Big(\frac{t^\cut}{\mu_B^2}\Big)^{\eta_B(\mu_B, \mu)}
\,.\end{align}
Analogous expressions hold for the hard and soft functions. The corresponding evolution factors $U_H$ and $U_S$ are given in \app{RG} together with the functions $K_i^g$ and $\eta_i^g$. For simplicity, we suppress the $R$-dependence of the $U_i$, which starts at NNLL and enters through the noncusp anomalous dimensions.

Writing out the evolution factors explicitly, the resummed cross section with a veto on $\Tau_{B,C}^\jet$ reads
\begin{align} \label{eq:TBCResum}
&\frac{\df\sigma_0^\resum}{\df Y}(\Tau_{B,C}^\jet \!< \Tau^\cut)
\nn \\ &\quad
= \sigma_B H_{gg}(m_t, m_H^2, \mu_H)\, B_g(m_H \Tau^{\cut},x_a,R,\mu_B)
\nn\\  &\quad\qquad \times
B_g(m_H \Tau^{\cut},x_b,R,\mu_B)\, S_{gg}^{B,C}(\Tau^{\cut},R, \mu_S)
\nn\\ &\quad\qquad \times
U_0(m_H,\Tau^{\cut}, \mu_H,\mu_B,\mu_S)
\nn \\ &\qquad
+ \frac{\df\sigma_0^{\text{Rsub}}}{\df Y}(\Tau_{B,C}^\jet \!< \Tau^\cut, R)
\,,\end{align}
and for a veto on $\Tau_{B\cm,C\cm}^\jet$, we have
\begin{align} \label{eq:TBCcmResum}
&\frac{\df\sigma_0^\resum}{\df Y}(\Tau_{B\cm,C\cm}^\jet \!< \Tau^\cut)
\nn \\ &\quad
= \sigma_B H_{gg}(m_t, m_H^2, \mu_H)\,B_g(m_H \Tau^{\cut}e^Y\!,x_a,R,\mu_B)
\nn\\  &\qquad\quad \times
B_g(m_H \Tau^{\cut}e^{-Y}\!,x_b,R,\mu_B)\, S_{gg}^{B,C}(\Tau^{\cut},R, \mu_S)
\nn\\ &\qquad\quad \times
U_0(m_H,\Tau^{\cut}, \mu_H,\mu_B,\mu_S)
\nn \\ &\qquad
+ \frac{\df\sigma_0^{\text{Rsub}}}{\df Y}(\Tau_{B\cm,C\cm}^\jet \!< \Tau^\cut, R)
\,.\end{align}
Here, the total evolution factor, combining the individual hard, beam, and soft ones, is
\begin{align} \label{eq:Utot}
&U_0(m_H,\Tau^{\cut}, \mu_H,\mu_B,\mu_S)
\\ &\quad
= U_H(m_H^2, \mu_H,\mu)\, U_B^2(m_H \Tau^\cut\!,\mu_B,\mu)\, U_S(\Tau^{\cut}\!,\mu_S,\mu)
\,.\nn\end{align}
The dependence on the common arbitrary scale $\mu$ cancels exactly between the individual $U_i$ due to RG consistency.
Note that the $U_{\rm tot}$ is the same in \eqs{TBCResum}{TBCcmResum}, because according to \eq{evolveB}
\begin{align}
&U_B(m_H \Tau^\cut e^Y\!,\mu_B,\mu)\, U_B(m_H \Tau^\cut e^{-Y}\!\!,\mu_B,\mu)
\nn \\ & \qquad
= U_B^2(m_H \Tau^\cut,\mu_B,\mu)
\,.\end{align}
Hence, the only difference between the $\Tau_{B,C}^\jet$ and $\Tau_{B\cm,C\cm}^\jet$ cross sections is the $Y$-dependence in the arguments of the fixed-order beam functions in \eqs{TBCResum}{TBCcmResum}.

\subsubsection{Ingredients at NLL$'$}

The resummation at the NLL$^\prime$ level includes the NLL RG evolution and in addition the fixed order one-loop expressions for the hard, beam and soft functions. (The latter provide the exact $\ord{\alpha_s}$ boundary conditions for the RGEs, which are formally a NNLL effect, but are important for matching to the full NLO cross section. This primed counting also has a number of other advantages and is frequently adopted, see e.g. \mycite{Ligeti:2008ac, Abbate:2010xh, Berger:2010xi, Alioli:2012fc, Liu:2012sz, Stewart:2013faa, Hoang:2014wka, Almeida:2014uva}.)

The $\ord{\alpha_s}$ hard function can be taken directly from \mycite{Berger:2010xi} and is given for completeness in \app{hard}. The beam functions can be computed as a convolution between perturbative matching kernels and the standard parton distribution functions (PDFs), $f_j$, as~\cite{Fleming:2006cd, Stewart:2009yx, Stewart:2010qs}
\begin{align}  \label{eq:BOPE}
B_{i}(t^{\cut}\!,x,R,\mu_B) &= \sum_j \!\int^1_x \!\! \dfrac{\df z}{z}\, \cI_{ij}(t^{\cut}\!,z,R,\mu_B) f_{j}\Bigl(\frac{x}{z},\mu_B \Bigr)
\nn\\ &\quad \times
\biggl[1 + \ORD{\frac{\lqcd^2}{t^{\cut}}} \biggr]
\,.\end{align}
At fixed $\ord{\alpha_s}$, the $B_{i}(t^{\cut}\!,x,R,\mu_B)$ can be obtained by integrating the one-loop differential $t$-dependent beam function~\cite{Stewart:2010qs, Berger:2010xi}, because the $\Tau_{B,C}^\jet$ measurement function for a veto on the emission of only one gluon is simply a theta function, $\theta(t<t^{\cut})$, of the virtuality $t$. Therefore, the one-loop gluon matching coefficient is
\begin{align}
 \cI_{gj}^{(1)}(t^{\cut},z,R,\mu_B) = \int_0^{t^{\cut}} \!\! \df t\; \cI_{gj}^{(1)}(t,z,\mu_B)
 \,,\end{align}
which is given in \app{beam}. We stress that this only holds for the one-loop fixed-order contributions. The resummed beam function in \eq{Bresum} is different from integrating the resummed differential $t$-dependent beam function already at NLL due to the different renormalization structure.
At two loops (and beyond) an explicitly $R$-dependent jet clustering correction term must be added to the integrated bare $t$-dependent beam functions~\cite{Gaunt:2014xga, Gaunt:2014cfa} to obtain the correct bare results for the $\Tau_{B,C}^\jet$-type observables. Since these $R$-dependent jet clustering corrections affect the UV divergences, the two-loop anomalous dimension of the beam function as well as its NNLL evolution explicitly depend on $R$.

Similarly, the one-loop soft function for $\Tau_{B}^\jet\!<\!\Tau^\cut$ can be obtained by integrating the one-loop soft function differential in beam thrust~\cite{Berger:2010xi}, see \app{softTauB}. The one-loop soft function for $\Tau_{C}^\jet<\Tau^\cut$ is explicitly calculated in \app{softTauC}. (It is directly related to the integrated one-loop soft function for the $C$-parameter event shape in $e^+e^-$ collisions.)

For the RG evolution at NLL, we require the cusp anomalous dimension to two loops~\cite{Korchemsky:1987wg}, and the noncusp anomalous dimensions to one loop. The one-loop coefficients of the noncusp anomalous dimensions in \eq{anomdimsstructure} are the same as for the corresponding beam thrust results~\cite{Berger:2010xi}. To see this, note that at $\ord{\alpha_s}$
\begin{align}
 \gamma_B^{g(1)}= -\mu \frac{\df}{\df \mu} Z_B^{g(1)}\,,\quad  \gamma_S^{g(1)}= -\mu \frac{\df}{\df \mu} Z_S^{g(1)}
\,,\end{align}
and the one-loop $\msb$ counterterms, $Z_i^{g(1)}$, for $\Tau_{B,C}^\jet\!<\!\Tau^\cut$ and inclusive beam thrust are also simply related by integration.

\subsection{Nonsingular contributions}
\label{nonsing}

In the previous section we have discussed the ingredients to the resummed part of the $\Tau^\jet_f$-veto cross section to NLL$'$ order.
To incorporate the full $\ord{\alpha_s}$ corrections of the fixed-order (FO) cross section at NLO, we must add the $\ord{\alpha_s}$ nonsingular contribution in \eq{fullXsec}, which is particularly relevant for large $\Tau^\cut$.

The FO nonsingular contribution differential in $\Tau^\jet_f$
is defined by the difference of the differential FO result in full QCD and the corresponding FO singular contribution,
\begin{equation}
\frac{\df\sigma_0^\nons}{\df\Tau_f^\jet \df Y}
= \frac{\df\sigma_0^\FO}{\df\Tau_f^\jet \df Y} - \frac{\df\sigma_0^\sing}{\df\Tau_f^\jet \df Y}
\,.\end{equation}
The FO singular terms in turn are given by a strict expansion of the resummed part of the cross section to a given fixed order in $\alpha_s(\mu_\FO)$, where $\mu_\FO \equiv \mu_r = \mu_f \sim m_H$ is the renormalization and factorization scale of the FO cross section. Suppressing the dependence on $\Tau^\jet_f$ and $Y$, we thus have
\begin{align}
 \df \sigma_0^\sing(\mu_\FO) &\equiv \df \sigma_0^\resum(\mu_H, \mu_B, \mu_S) \Big|_{\FO \text{ in } \alpha_s(\mu_\FO)} \\
 &= \df \sigma_0^\resum(\mu_H = \mu_B = \mu_S = \mu_\FO)
\label{eq:singdef}
\,.\end{align}
A priori, \eq{singdef} is only true up to higher orders in $\alpha_s(\mu_\FO)$. However, we always reexpand the product of fixed-order contributions to the hard, beam, and soft functions entering in $\df \sigma_0^\resum$ in \eqs{TBCResum}{TBCcmResum}, such that \eq{singdef} holds exactly. At NLL$'$ this means that $\df \sigma_0^{\resum} \propto H^{(0)}B^{(0)}B^{(0)}S^{(0)} + H^{(1)} B^{(0)}B^{(0)}S^{(0)} + H^{(0)}B^{(1)}B^{(0)}S^{(0)} + \ldots\,$, where the superscripts $(0)$ and $(1)$ indicate the LO and NLO fixed-order contributions to $H(\mu_H)$, $B(\mu_B)$, and $S(\mu_S)$, respectively.
In this way we ensure that when turning off the resummation in the NLL$'$ result by setting $\mu_H = \mu_B = \mu_S =\mu_\FO$, we exactly reproduce the NLO cross section
\begin{align} \label{eq:nonsing}
\frac{\df \sigma_0^\FO}{\df \Tau_f^\jet \df Y}(\mu_\FO)  &= \frac{\df \sigma_0^{\resum}}{\df \Tau_f^\jet \df Y}(\mu_H = \mu_B = \mu_S = \mu_\FO)  \nn\\
& \quad + \frac{\df \sigma_0^\nons}{\df \Tau_f^\jet \df Y}(\mu_\FO)
\,, \end{align}
and the analogous relation holds for the jet-vetoed cross section integrated over $\Tau_f^\jet < \Tau^\cut$.

In practice, we use \eq{nonsing} to determine the nonsingular corrections. The resummed result differential in $\Tau_f^\jet$ and evaluated at $\mu_\FO$ can be obtained by taking the derivative of the resummed cumulant cross sections in \eqs{TBCResum}{TBCcmResum} with respect to $\Tau^\cut$. (Alternatively the FO singular result can be obtained by directly combining the differential FO expressions for the hard, beam, and soft functions evaluated at the common scale $\mu_i = \mu_\FO$.)
Furthermore, we have calculated the relevant differential $\ord{\alpha_s}$ cross sections for all four observables $\Tau^\jet_f$ in full QCD, allowing us to determine the nonsingular contributions via \eq{nonsing}. The details of these calculations will be presented elsewhere~\cite{ournonsing}. The results for the full, singular, and nonsingular cross sections at NLO differential in $\Tau_f^\jet$ (and integrated over the full $Y$-range) are shown in \fig{Singnonsing} as a function of $\Tau^\jet_f$.
By construction, the NLO nonsingular terms differential in $\Tau_f^\jet$ only have integrable singularities for $\Tau_f^\jet \to 0$, so we can integrate them over $\Tau^\jet_f < \Tau^\cut$ to obtain the nonsingular cumulant $\df\sigma^\nons(\Tau_f^\jet < \Tau^\cut, \mu_\FO)$.

The resummed contribution is evolved to an imaginary hard scale, which avoids large corrections in the hard function when evaluated at a timelike argument $q^2 = m_H^2$~\cite{Parisi:1979xd, Sterman:1986aj, Magnea:1990zb, Ahrens:2008qu}. For consistency, we have to include the same evolution also in the nonsingular contributions~\cite{Berger:2010xi, Stewart:2013faa}, which at NLO simply amounts to multiplying it by the hard evolution factor. The final NLO nonsingular contribution is then given by
\begin{align} \label{eq:fullnons}
\frac{\df\sigma_0^\nons}{\df Y}(\Tau_f^\jet \!< \Tau^\cut)
&= \frac{H_{gg}^{(0)}(-\img \mu_{\rm ns})}{H_{gg}^{(0)}(\mu_{\rm ns})} U_H(-\img\mu_{\rm ns}, \mu_{\rm ns})
\nn\\ & \quad\times
\frac{\df\sigma_0^{\nons(1)}}{\df Y}(\Tau_f^\jet \!< \Tau^\cut, \mu_{\rm ns})
\,,\end{align}
where we introduced $\mu_{\rm ns}$ to denote the scale at which the nonsingular contributions are evaluated.
Combined with the resummed contribution according to \eq{fullXsec}, this yields the complete cross section for a $\Tau_f^\jet$ veto at NLL$'+$NLO.

\subsection{Scale choices}
\label{subsec:scalevar}

\begin{figure*}[t!]
\begin{center}
\includegraphics[width=0.33\textwidth]{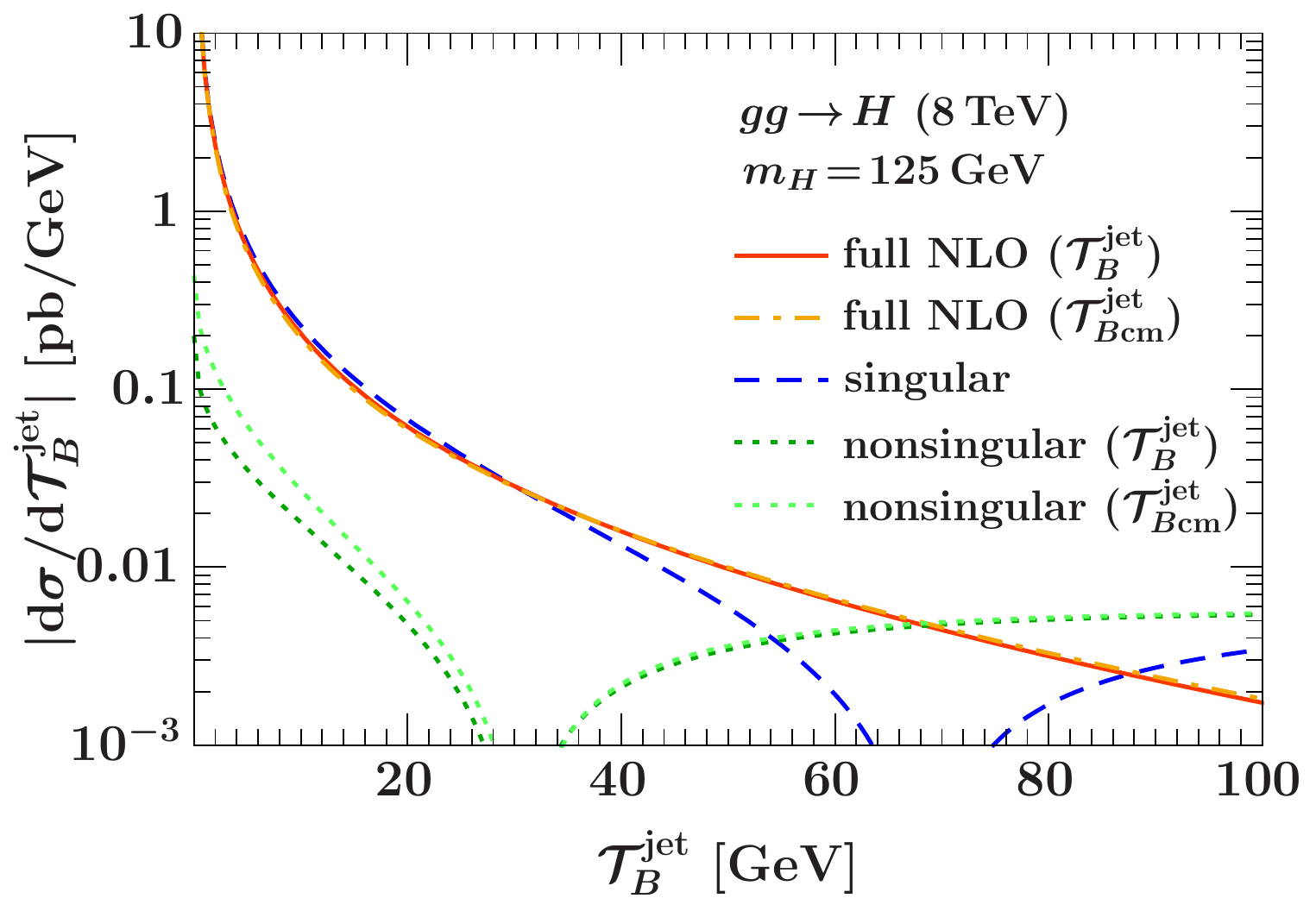}%
\includegraphics[width=0.33\textwidth]{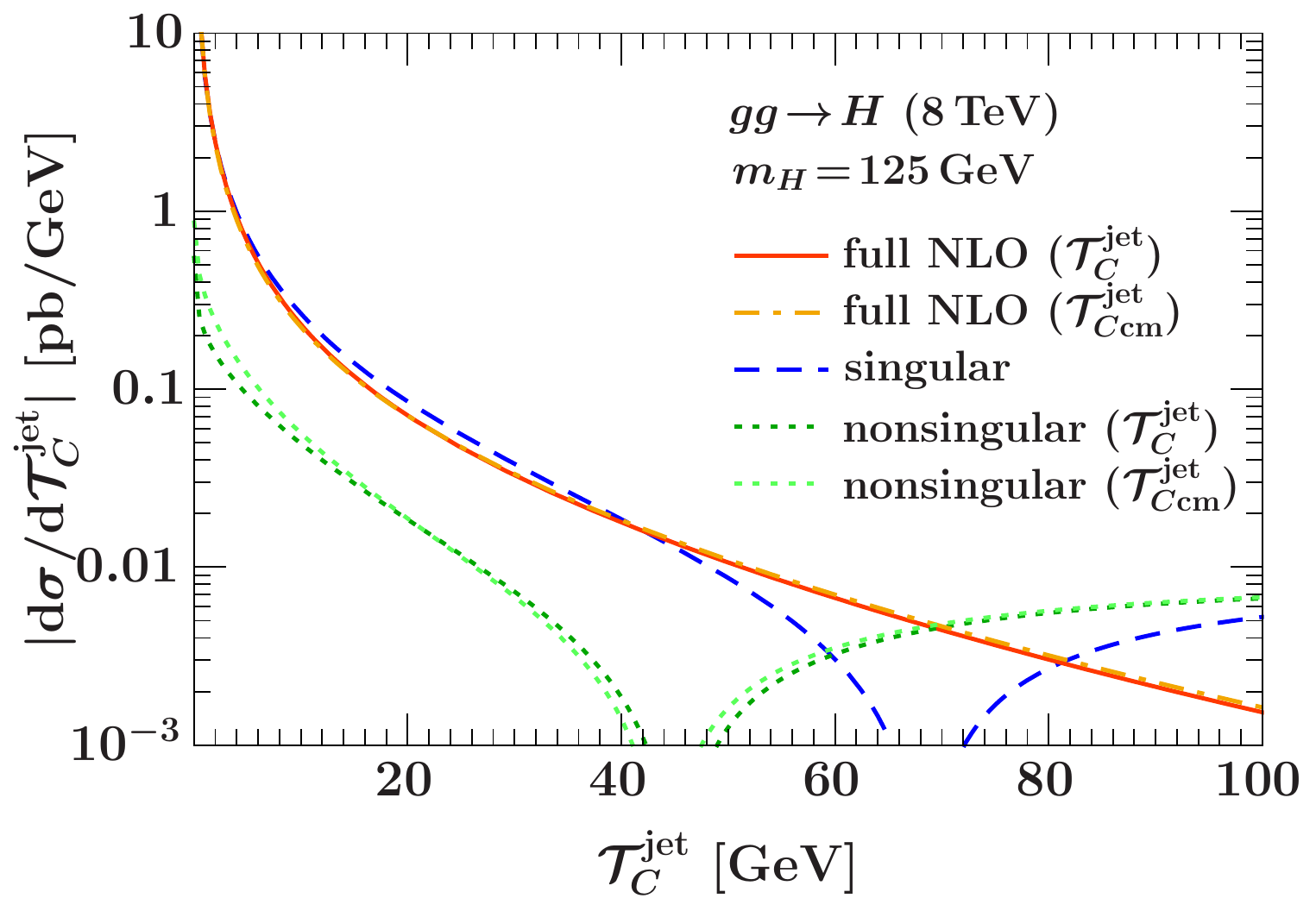}%
\hfill%
\includegraphics[width=0.315\textwidth]{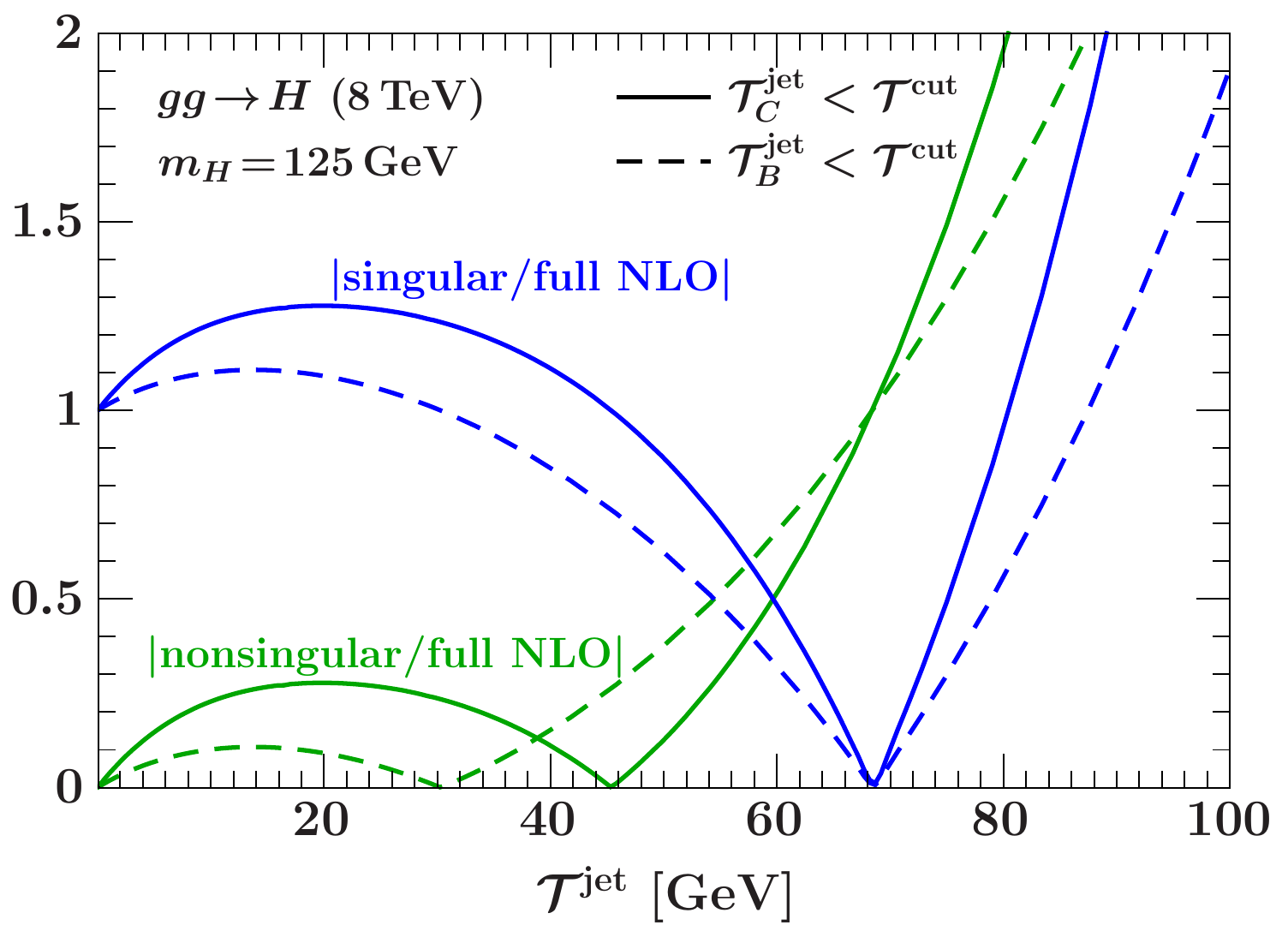}%
\end{center}
\vspace{-1ex}
\caption{Comparison of the singular, nonsingular, and full NLO cross sections differential in $\Tau_f^\jet$ and integrated over all of $Y$. The left and middle plots show the magnitude of the differential cross sections for $\Tau_B^\jet$ and $\Tau_C^\jet$ on a logarithmic scale. The right plot shows the ratios of nonsingular and singular contributions to the full NLO cross section for both $\Tau_B^\jet$ and $\Tau_C^\jet$.}
\label{fig:Singnonsing}
\end{figure*}

In this subsection, we discuss how to choose appropriate beam and soft scales as a function of $\Tau^\cut$.
For this purpose, we have to compare the relative size of the singular and nonsingular contributions in relation to the full FO cross section in different regions of $\Tau^\jet_f$. For this comparison, we integrate over the full $Y$-range. The left and middle panels of \fig{Singnonsing} show the magnitude of the differential singular, nonsingular, and full FO cross sections for $\Tau_{B(\cm)}^\jet$ and $\Tau_{C(\cm)}^\jet$. The curves for $\Tau_{B\cm}^\jet$, $\Tau_{C\cm}^\jet$ are displayed in light colors and for $\Tau_B^\jet$, $\Tau_C^\jet$ in darker colors. In the right panel of \fig{Singnonsing}, we plot the magnitude of the ratio of the singular and nonsingular contributions to the full NLO cross section for both $\Tau_B^\jet$ and $\Tau_C^\jet$.

Note that the singular differential contribution is identical for all four $\Tau_f^\jet$ variables, because the difference between $\Tau_{B(\cm)}^\jet$ and $\Tau_{C(\cm)}^\jet$ only appears as a constant ($\Tau_f^\jet$-independent) term in the soft function and does not affect the singular spectrum.
Also the (explicit) total $Y$ dependence resides in the $\Tau^\cut$-independent part of the NLO singular contribution of the $\Tau_{B\cm}^\jet$ and $\Tau_{C\cm}^\jet$ vetoed cross section and drops out in the spectrum [which can be seen from \eq{TBCcmResum} together with \eqs{Igg}{Igq}].
The full FO cross section, however, depends on the specific measurement function and is different for all four observables. 
We therefore observe significant differences when comparing the nonsingular contributions for $\Tau_B^\jet$ and $\Tau_C^\jet$ in \fig{Singnonsing}.
The nonsingular contributions for $\Tau_{B\cm}^\jet$ and $\Tau_{C\cm}^\jet$ (light green dotted lines) are slightly larger than the corresponding ones for $\Tau_B^\jet$ and $\Tau_C^\jet$.
This is due to $Y$-dependent terms that are not captured by the resummed singular contributions and are thus part of the nonsingular contributions in \eq{nonsing}.

The plots show that the nonsingular contributions are power suppressed for small values of $\Tau_f^\jet$, become comparable to the singular contributions around $\Tau_f^\jet \sim 60\GeV$, and exceed the FO cross section beyond $\Tau_f^\jet \gtrsim 70\GeV$. Based on these observations, we can distinguish three different regions according to the relative size of the singular and nonsingular contribution to the full FO cross section at increasing $\Tau_f^\jet$: resummation, transition, and fixed-order regions. (In principle, there is a fourth nonperturbative regime $\Tau_f^\jet \lesssim \lqcd$, which is however not relevant for our discussion.)

In the resummation region, i.e. at low values $\Tau_f^\jet \ll m_H$, the singular contribution dominates and must be resummed while the nonsingular contributions are power corrections suppressed by $\Tau_f^\jet/m_H$. To correctly resum the large logarithms in this region, the scales should follow their canonical relations as dictated by the factorization of the jet-vetoed cross section,
\begin{align} \label{eq:canonical}
\mu_H = -\img m_H \,,\quad \mu_B = \sqrt{m_H \mu_S} \,, \quad
\mu_S = \Tau^\cut
\,.\end{align}
At large values of $\Tau^\cut \gtrsim m_H/2$, the singular and nonsingular contributions are equally important and there are large cancellations between the two, which would be spoiled if the resummation is kept on. Hence, in this region the resummation must be turned off, which is achieved by letting all scales approach a common FO scale,
\begin{align}
\abs{\mu_H} = \mu_B = \mu_S = \mu_{\rm ns} = \mu_\FO
\,,\end{align}
which then ensures that the result correctly reproduces the total cross section [cf. \eqs{nonsing}{fullnons}]. By keeping the hard scale at an imaginary value, this becomes the $\pi^2$-improved FO cross section, which exhibits an improved perturbative convergence.

In the transition between the resummation and fixed-order regions, both the resummed logarithmic corrections as well as the nonsingular FO contributions are numerically important.
To optimally describe this region, which is often also the experimentally most relevant one, we employ profile scales~\cite{Ligeti:2008ac, Abbate:2010xh} that incorporate the constraints in \eq{canonical}, towards small values of $\Tau_f^\jet$ and provide a smooth interpolation to $\mu_\FO$ at large values of $\Tau_f^\jet$. For our choice of profile scales and the related estimation of perturbative uncertainties we adapt the discussion of the $p_T^\jet$-veto in \mycite{Stewart:2013faa} to our present case, where we have virtuality-like (SCET-I) as opposed to $p_T$-like (SCET-II) scale relations.

For the central profiles we take
\begin{align} \label{eq:centralscale}
\mu_H &= -\img \mu_\FO
\,, \nn \\
\mu_S(\Tau^\cut) & = \mu_\FO f_\run(\Tau^\cut/m_H)
\,, \nn \\
\mu_B(\Tau^\cut) &= \sqrt{\mu_S(\Tau^\cut) \mu_\FO} = \mu_\FO \sqrt{f_\run(\Tau^\cut/m_H)}
\,, \nn \\
\mu_{\rm ns} &= \mu_\FO
\,,\end{align}
where the common profile function $f_\run(x)$ is as in \mycite{Stewart:2013faa},
\begin{align}
f_{\run}(x) &= 
\begin{cases} x_0 \big(1+\frac{x}{4 x_0}\big) & x \le 2x_0 
 \\ x & 2x_0 \le x \le x_1
 \\ x + \frac{(2-x_2-x_3)(x-x_1)^2}{2(x_2-x_1)(x_3-x_1)} & x_1 \le x \le x_2
 \\  1 - \frac{(2-x_1-x_2)(x-x_3)^2}{2(x_3-x_1)(x_3-x_2)} & x_2 \le x \le x_3
 \\ 1 & x_3 \le x
\end{cases}\,.
\label{eq:frun}
\end{align}
The first regime in \eq{frun} for $x \le 2x_0$ is the nonperturbative regime, where we let the scales $\mu_B$ and $\mu_S$ approach fixed values $\sqrt{x_0}\mu_\FO > \Lambda_{\rm QCD}$ and $x_0 \mu_\FO > \Lambda_{\rm QCD}$ respectively as $x \rightarrow 0$.
For $x \sim \lqcd/m_H$, corresponding to $\Tau_f^\jet\sim\lqcd$, our purely perturbative predictions are insufficient to correctly describe the cross sections, since here nonperturbative corrections can become of $\ord{1}$. In practice, this region is irrelevant and has no effect on the cumulant jet-vetoed cross sections that we are interested in.

The second line in \eq{frun} corresponds to the resummation region and yields the canonical scaling in \eq{canonical}.
The third and fourth lines describe the transition region. They provide a quadratic scaling for a smooth transition to the FO region (last line), where all the scales are equal and the resummation is turned off.

To fix the profile parameters $x_i$ in \eq{frun} we first choose a value for $x_3$, where the resummation is turned off completely. This should happen roughly after the point, where the singular spectrum vanishes (the singular cumulant has a maximum) so the nonsingular spectrum is equal to the full result. In addition, it should certainly happen before the point, where the singular spectrum has the same magnitude but opposite sign as the full and the nonsingular becomes twice the size of the full result, since at this point there is clearly an $\ord{1}$ cancellation between singular and nonsingular. Hence, for $\TauB^\jet$ we choose $x_3 = 0.6$ corresponding to $\TauB^\jet = 75\GeV$. For $\TauC^\jet$ we choose $x_3 = 0.55$, since here the singular-nonsingular cancellations set in a bit earlier. For $x = \Tau_f^\jet/m_H \lesssim 0.1$ the physical scales are separated by an order of magnitude (and the nonsingular are suppressed by an order of magnitude). Hence, a natural choice for $x_1$ is of $\ord{0.1}$. We use $x_1 = 0.15$ for $\TauB^\jet$ and $x_1 = 0.1$ for $\TauC^\jet$, which ensures that the size of the transition region, $x_3 - x_1$, is the same for both and also long enough for the scales to smoothly transit to the FO region. The midpoint of the transition region, $x_2$, is then fixed by setting $x_2=(x_3 - x_1)/2$. Note that although the strict canonical scaling stops at $x_1$, the resummation is still important all the way through the transition region, at least until $x_2$, and starts to get turned off beyond. To summarize, our central profile parameters for $\Tau_{B(\cm)}^\jet$ are
\begin{align} \label{eq:TauBprofile}
\mu_\FO &= m_H
\,, \qquad
x_0 = 2.5\GeV/\mu_\FO
\,, \nn \\
\{x_1,x_2, x_3\} &= \{0.15,0.375,0.6\}
\,,\end{align}
and for $\Tau_{C(\cm)}^\jet$ they are
\begin{align} \label{eq:TauCprofile}
\mu_\FO &= m_H
\,, \qquad
x_0 = 2.5\GeV/\mu_\FO
\,, \nn \\
\{x_1,x_2, x_3\} &= \{0.1,0.325,0.55\}
\,.\end{align}
The resulting central profiles for $\mu_B$ and $\mu_S$ are shown in the middle and right panels of \fig{scaleunc} by the green and orange solid lines.

\subsection {Fixed-order and resummation uncertainties}
\label{subsec:uncertainties}

\begin{figure*}[t!]
\begin{center}
\includegraphics[width=0.33\textwidth]{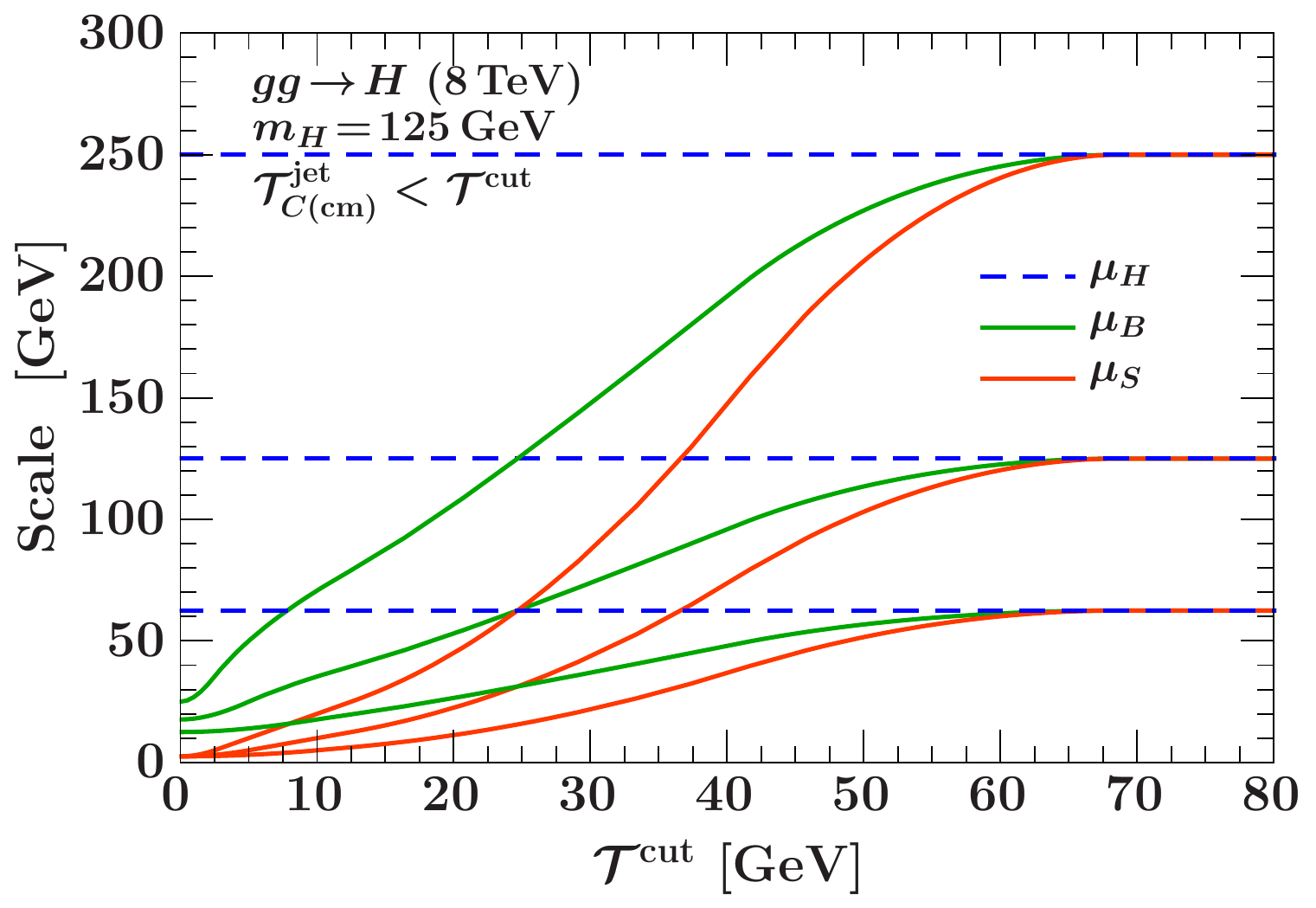}%
\hfill%
\includegraphics[width=0.33\textwidth]{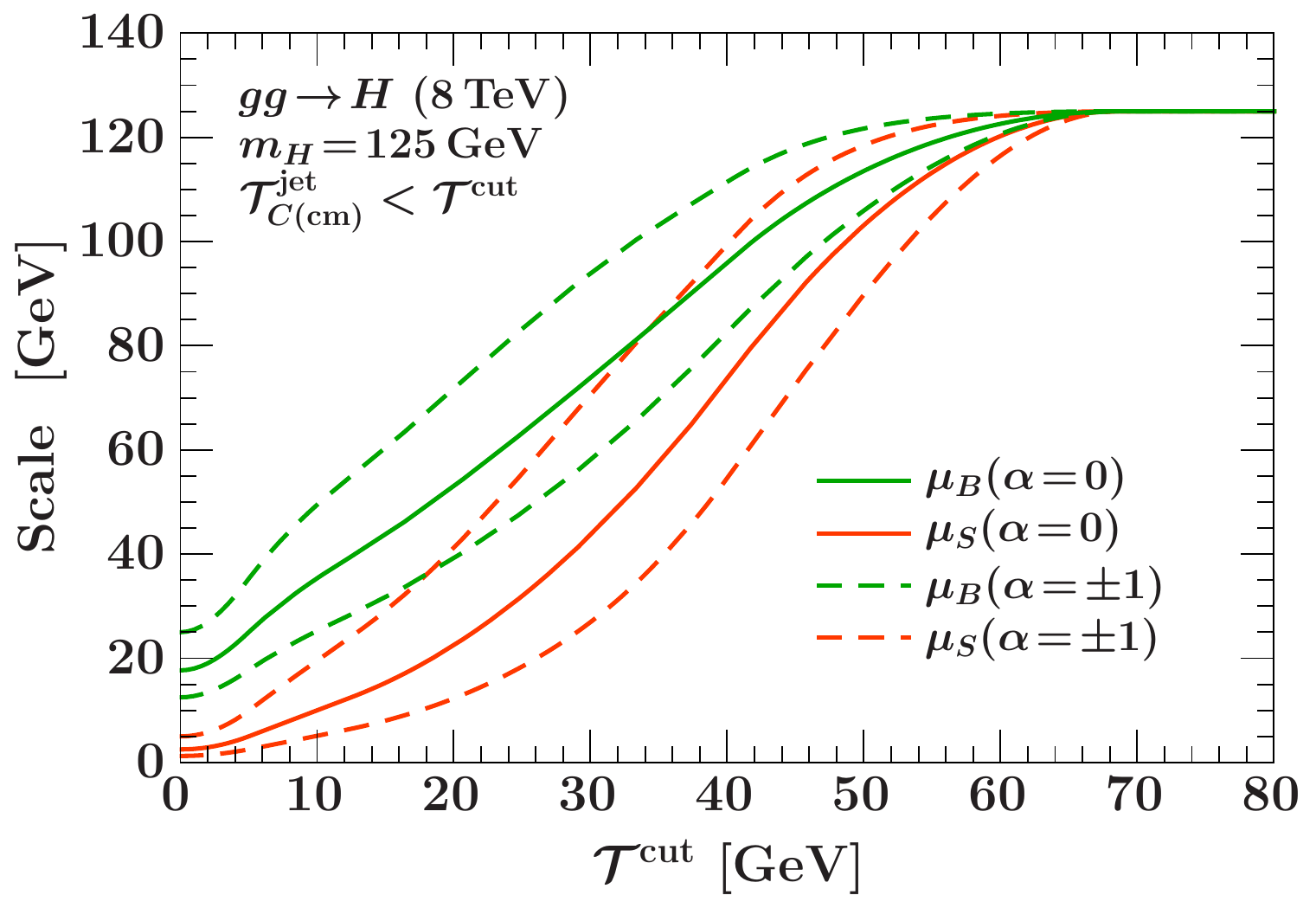}%
\hfill%
\includegraphics[width=0.33\textwidth]{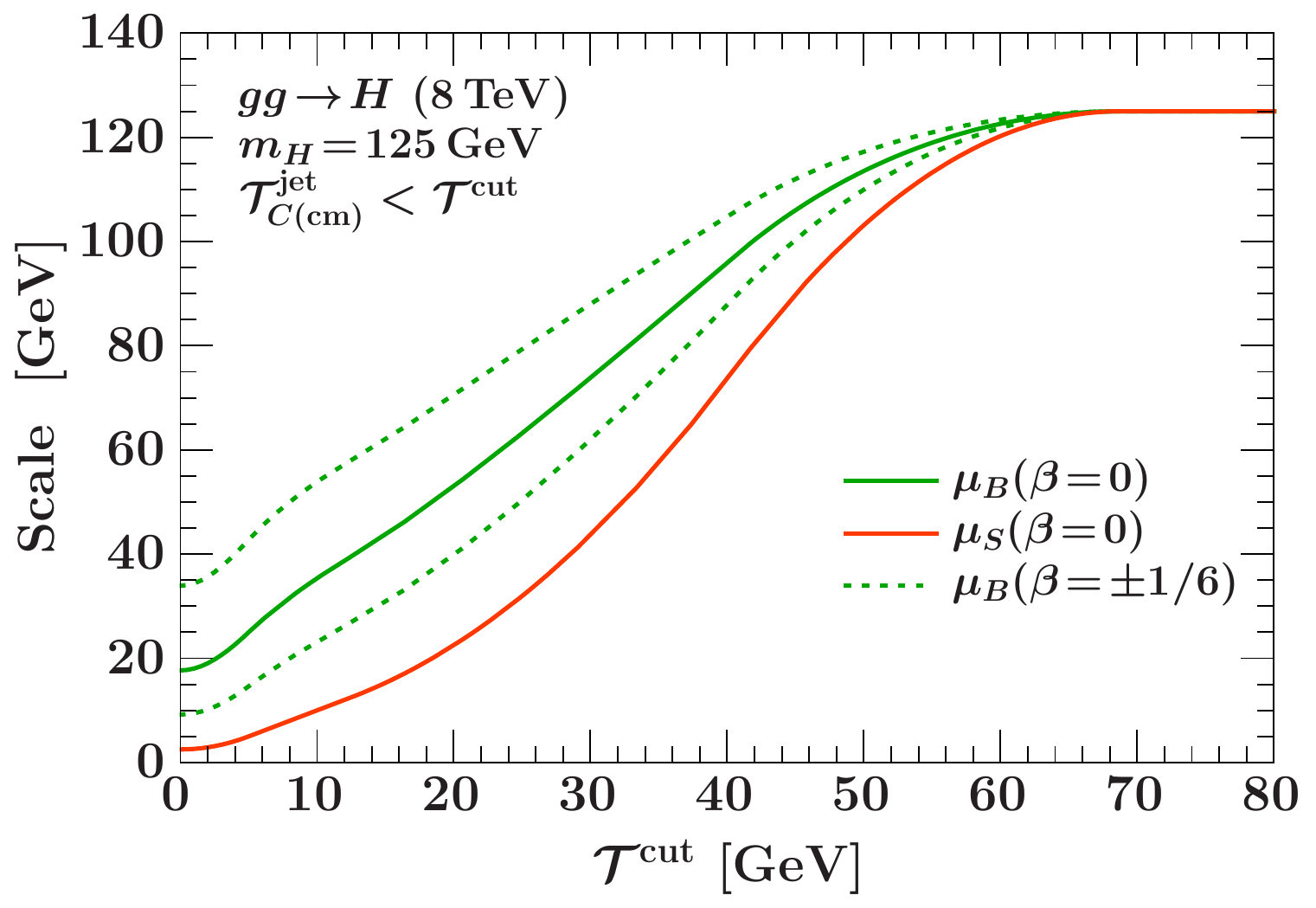}%
\hfill%
\end{center}
\vspace{-1ex}
\caption{Profile scale variations as described in the text. The left plot shows the collective variation of all scales by a factor of two, which is used to estimate the FO uncertainty. The middle and right plots shows the variations of the beam and soft scales used to estimate the resummation uncertainty.}
\label{fig:scaleunc}
\end{figure*}

A key aspect of precision cross section predictions is to reliably estimate the perturbative uncertainties.
A convenient and physically motivated way to parametrize the theoretical uncertainties in jet-vetoed cross sections is in terms of fully correlated (yield) and fully anticorrelated (migration) components~\cite{Berger:2010xi, Stewart:2011cf, Gangal:2013nxa, Stewart:2013faa}, and we follow the same logic here.

Within our resummation framework, these two components are naturally associated with two distinct types of uncertainties, see \mycites{Stewart:2011cf, Stewart:2013faa} for a detailed discussion. First, the resummation uncertainty $\Delta_\resum$ corresponds to the intrinsic uncertainty in the resummed logarithmic series induced by the jet veto (or jet binning) cut. It must be anticorrelated between the cross section that survives the jet veto (the $0$-jet bin) and the cross section that is vetoed (the $\geq 1$-jet bin), such that it cancels in the total inclusive cross section given by their sum. Hence, we can identify $\Delta_\resum$ with the migration uncertainty.
Second, the fixed-order uncertainty, $\Delta_{\FO}$, comes from scale variations in the FO contributions of the full resummed cross section, such that for large $\Tau^\cut$ it reproduces the FO scale variation uncertainty of the total cross section. It is identified with the yield uncertainty, and effectively probes the size of higher-order nonlogarithmic terms at any value of $\Tau^\cut$. Note that despite its naming, at small $\Tau^\cut$ it does so \emph{within} the resummed prediction.
The total uncertainty in the Higgs$+0$-jet cross section is then given by
\begin{align} \label{eq:totunc}
\Delta_0^2(\Tau^\cut) = \Delta_\FO^2(\Tau^\cut) + \Delta_\resum^2(\Tau^\cut)
\,.\end{align}

To evaluate $\Delta_\FO$, we take the collective variation of all scales $\mu_i$ up and down by a factor of 2, as shown in the left panel of \fig{scaleunc}. This is done by setting $\mu_\FO =\{2m_H,m_H/2\}$ in \eq{centralscale}.
At large $\Tau^\cut$ values, this yields the standard scale variation of the ($\pi^2$-improved) FO cross section.
By varying $\mu_\FO$, all the ratios between the scales $\mu_H$, $\mu_B$, and $\mu_S$ are kept fixed, so that the arguments of the logarithms that are resummed in the evolution factors $U_i$ remain unchanged.
We stress that the scales do not represent physical input quantities. Rather, the changes observed in the cross section resulting from the scale variations are simply an indicator of the possible size of higher-order corrections. In particular, one should not attribute any meaning to possibly asymmetric up/down variations in the cross section. Instead, we take the maximal observed deviation from the central value as our perturbative uncertainty estimate. Thus, we adopt
\begin{align} \label{eq:DeltaFO}
\Delta_\FO (\Tau^\cut) = \max_{v \in V_\FO} \Abs{\sigma_0^v(\Tau^\cut) - \sigma_0^\central (\Tau^\cut)}
\end{align}
for the FO uncertainty in \eq{totunc}, where $V_\FO$ denotes the variations $\mu_\FO =  \{2m_H$, $m_H/2\}$.  

Next, to estimate the resummation uncertainty, $\Delta_\resum$, we vary the profile scales for $\mu_B$ and $\mu_S$ defined in the previous section about their central profile while keeping $\abs{\mu_H}=\mu_\FO = m_H$ fixed. The aim is to vary the logarithms in the resummation factors $U_i$, in order to estimate the potential size of higher-order corrections in the resummed logarithmic series. At the same time, the scales must retain the natural scale hierarchy in the resummation region (as obeyed by the central scales),
\begin{align}
\mu_\FO \sim \mu_H \gg \mu_B \sim \sqrt{\mu_H \mu_S} \gg \mu_S\,,
\label{eq:hierarchy}
\end{align}
for all variations. 

First, we define a multiplicative factor
\begin{align}
f_\vary(x) = \begin{cases}
2(1-x^2/x_3^2) & 0 \le x \le x_3/2
 \\
1 + 2(1-x/x_3)^2 & x_3/2 \le x \le x_3
 \\
1 & x_3 \le x
\end{cases}\,,
\end{align}
which approaches 2 for $\Tau^\cut \rightarrow 0$ and 1 for $\Tau^\cut \rightarrow x_3m_H$, where the resummation is turned off.
The up and down variations of $\mu_S$ are then parametrized as
\begin{align} \label{eq:muSvary}
\mu_S^\vary(x, \alpha) &= f_\vary^\alpha (x)\, \mu_S(x)  =  \mu_\FO\, f_\vary^\alpha (x) \,f_\run(x)
\,.\end{align}
For the $\mu_B$ variations we define
\begin{align} \label{eq:muBvary}
\mu_B^\vary(x, \alpha,\beta)
&= {\mu_S^\vary(x, \alpha)}^{1/2-\beta} \mu_\FO^{1/2+\beta}
\\ 
&= \mu_\FO \bigl[f_\vary^\alpha(x)\, f_\run(x) \bigr]^{1/2-\beta}
\,,\end{align}
where the parameter $\beta$ modifies the exact canonical relation of the beam and soft scales in \eq{canonical}, to allow for a variation of $\mu_B$ independent of $\mu_S$. The central scales in \eq{centralscale} correspond to setting $\alpha = \beta = 0$.
The $\mu_B$ and $\mu_S$ variations we will perform are illustrated in the middle and right panels of \fig{scaleunc}, and are discussed in detail in the following. Note that all $\mu_B$ and $\mu_S$ variations turn off at large $\Tau^\cut$ (beyond $x_3$), such that the resummation uncertainty vanishes by construction when the resummation itself is turned off.

The arguments of the logarithms resummed in the overall evolution factor, \eq{Utot}, are given by the ratios of the three scales $\mu_H$, $\mu_B$, and $\mu_S$. Because of cancellations due to RG consistency the two relevant independent scale ratios entering the resummed logarithms are
\begin{align} \label{eq:scaleratios}
\frac{\mu_B^2}{\mu_H^2} \sim \frac{\Tau^\cut}{m_H}
\,,\qquad
\frac{\mu_S^2}{\mu_B^2} \sim \frac{\Tau^\cut}{m_H}
\,.\end{align}
This can be seen best by setting the arbitrary common renormalization scale $\mu=\mu_B$, such that $U_B=1$ and we are left with only two independent evolution factors $U_H$ and $U_S$, which resum logarithms of the scale ratios in \eq{scaleratios}. (The third possible scale ratio $\mu_S/\mu_H \sim \Tau^\cut/m_H$ is not independent as it can never appear alone in the evolution.)

We use the same $\alpha$ for both $\mu_B$ and $\mu_S$, which ensures that we never violate the parametric scaling $\mu_B^2 \sim \mu_S\mu_H$ when changing $\mu_S$. Varying $\alpha$ while keeping $\beta$ fixed in this setup then induces equal changes to the logarithms of the scale ratios in \eq{scaleratios} of the form
\begin{align}
\ln \frac{\mu_B^2}{\mu_H^2} \rightarrow \ln f_\vary^\alpha + \ln \frac{\mu_B^2}{\mu_H^2}
\,,\nn \\
\ln \frac{\mu_S^2}{\mu_B^2} \rightarrow \ln f_\vary^\alpha + \ln \frac{\mu_S^2}{\mu_B^2}
\,.\end{align}
On the other hand, varying $\beta$ with $\alpha$ fixed induces changes in the logarithms of equal magnitude but opposite sign,
\begin{align}
\ln \frac{\mu_B^2}{\mu_H^2} \rightarrow (1 - 2\beta) \ln \frac{\mu_B^2}{\mu_H^2}
\,,\nn \\
\ln \frac{\mu_S^2}{\mu_B^2} \rightarrow (1 + 2\beta) \ln \frac{\mu_S^2}{\mu_B^2}
\,.\end{align}
Hence, separate variations of $\alpha$ and $\beta$ independently probe the resummation of both types of logarithms.
(Changing them together, would effectively double-count the variation for one or the other set of logarithms.)

\begin{figure*}[t!]
\begin{center}
\includegraphics[width=0.68\columnwidth]{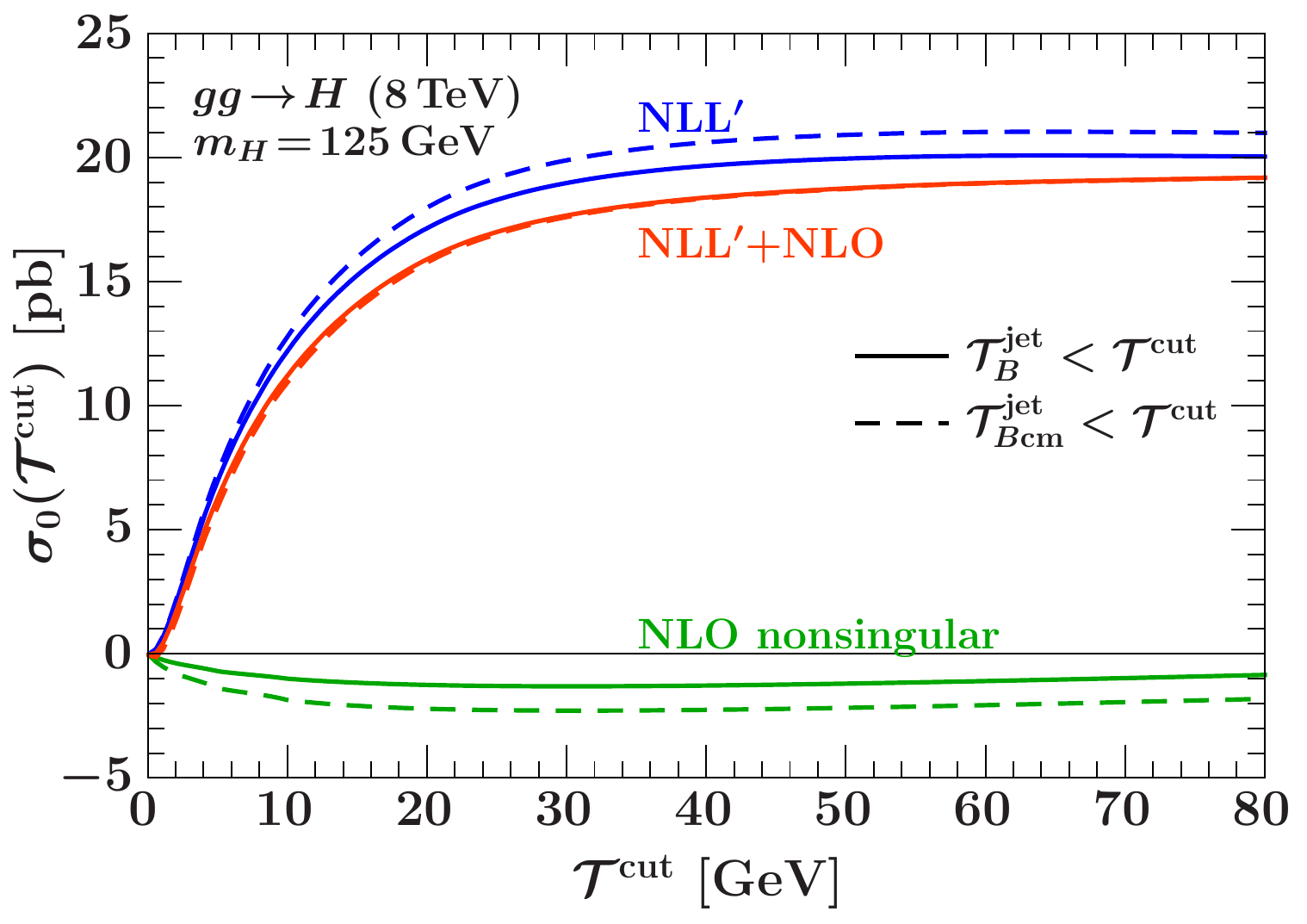}%
\hfill%
\includegraphics[width=0.68\columnwidth]{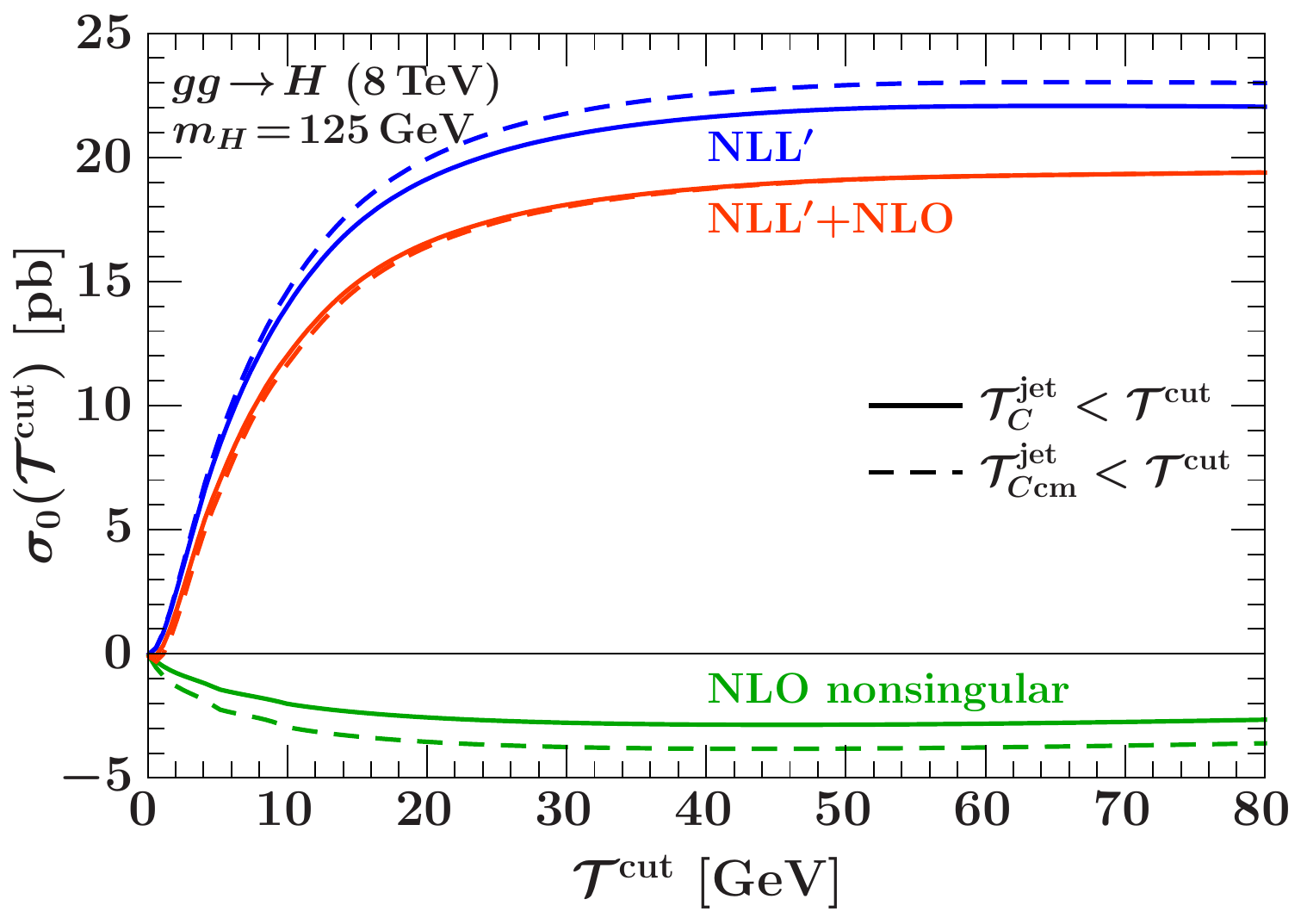}%
\hfill%
\includegraphics[width=0.68\columnwidth]{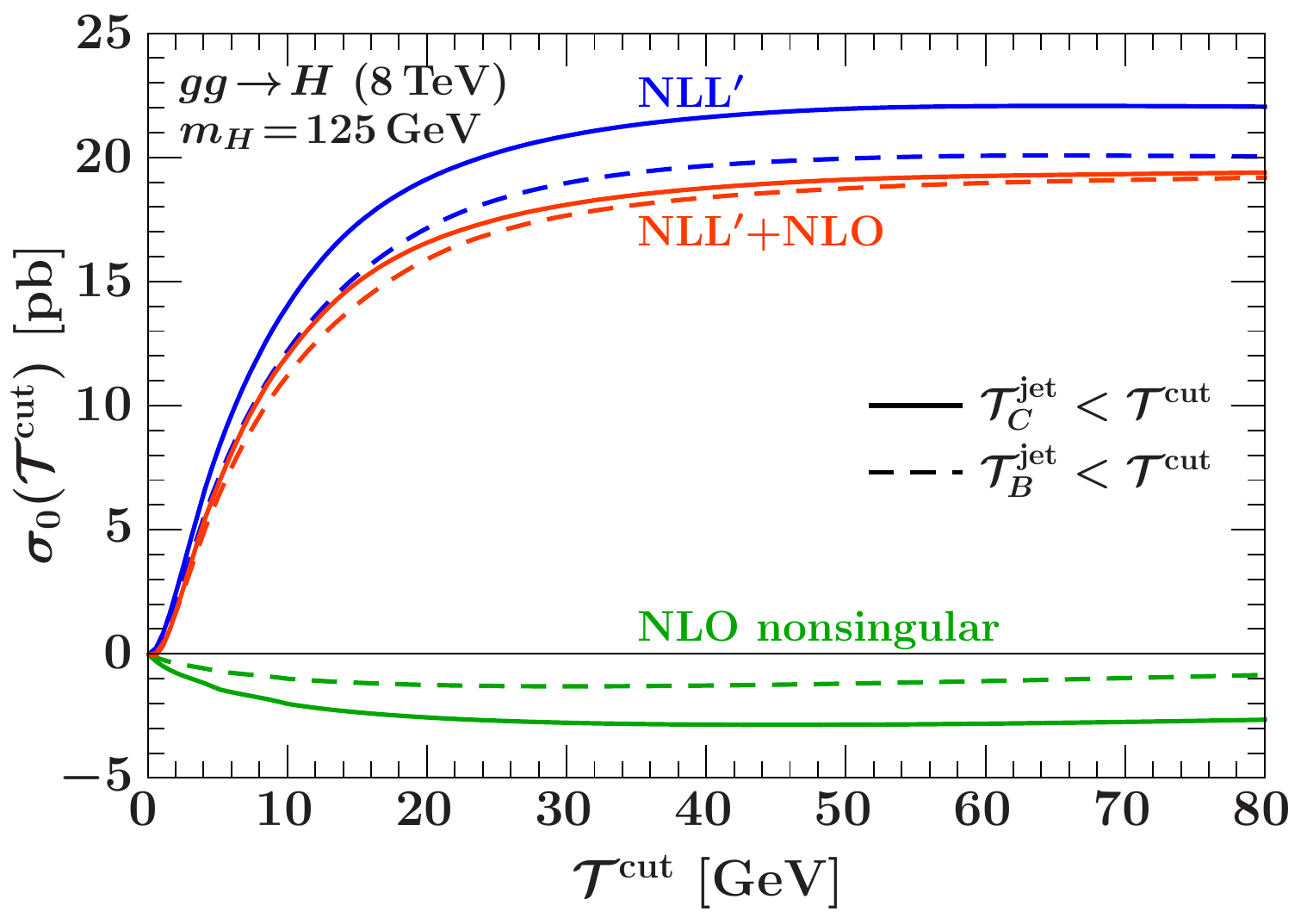}%
\end{center}
\vspace{-1ex}
\caption{Cumulant NLL$^\prime$ resummed and NLO nonsingular cross sections as a function of $\Tau^\cut$. The two plots on the left show the comparison between the two variants of $\TauB$-type and $\TauC$-type veto cross sections, respectively. The plot on the right shows the comparison between the corresponding contributions to $\TauC$ and $\TauB$-veto cross sections. The resummation/FO scales in the cross sections displayed here are given by the central profiles defined in \subsec{scalevar}.}
\label{fig:NLLpcentral}
\end{figure*}

The precise range of $\alpha$ and $\beta$ values is to some extent arbitrary. For our analysis of the resummation uncertainty $\Delta_\resum$ we choose the four parameter sets
\begin{align} \label{eq:alphabeta}
(\alpha, \beta) &= \{ (+1, 0),\, (-1, 0),\, (0, -1/6),\, (0, +1/6) \}
\,,\end{align}
which fullfill the requirements in \eq{hierarchy}. The $\alpha$ variation is shown by the dashed curves in the middle panel of \fig{scaleunc}. It yields the typical factor of $2$ variation in the soft scale for $\Tau^\cut \to 0$, and a corresponding factor $\sqrt{2}$ in $\mu_B$. The $\beta$ variation modifies the canonical relation between $\mu_B$ and $\mu_S$ ``half-way'' from $\mu_B^2 = \mu_S^{1/3} \mu_H^{2/3}$ to $\mu_B^2 = \mu_S^{2/3} \mu_H^{1/3}$, and is shown by the dotted lines in the right panel of \fig{scaleunc}. For most of the relevant $\Tau^\cut$ range, all four variations have an effect of similar size on the scale ratios in \eq{scaleratios}. For $\Tau^\cut\to 0$, the $\beta$ variation generates roughly a factor of $2$ variation in $\mu_B$, while keeping $\mu_S$ fixed. (Since for small $\Tau^\cut$ the scales $\mu_H$, $\mu_B$, and $\mu_S$ are widely separated, this still maintains the required scale hierarchy.)

We then define the overall resummation uncertainty as the maximum absolute deviation from the cross section evaluated with central profiles when performing the $\mu_B$ and $\mu_S$ profile scale variations,
\begin{align} \label{eq:Deltaresum}
\Delta_\resum(\Tau^\cut) =\!\! \max_{v \in V_\resum}\!\!\Abs{\sigma_0^{v}(\Tau^\cut) - \sigma_0^\central(\Tau^\cut)}
\,,\end{align}
where $V_\resum$ denotes the set of four variations in \eq{alphabeta}. This resummation uncertainty together with the fixed-order uncertainty in \eq{DeltaFO} then determines the total uncertainty of the 0-jet cross section as given in \eq{totunc}.

Finally, we should mention that in principle one should also vary the other profile parameters $x_0$ and $\{x_1, x_2, x_3\}$ in \eqs{TauBprofile}{TauCprofile}. However, at the NLL$'+$NLO order we are working, the resulting cross section variations are much smaller than those from varying $\mu_\FO$, $\alpha$, $\beta$. This could change at higher orders, at which point these additional profile parameter variations should be included.

\section{Results}
\label{sec:results}

In this Section, we present numerical results for the Higgs$+$0-jet gluon-fusion cross sections, $\sigma_0(\Tau_f^\jet\!<\!\Tau^\cut)$, using the four rapidity-weighted observables $\Tau_f^\jet=\Tau_{B}^\jet, \Tau_{B\cm}^\jet, \Tau_{C}^\jet, \Tau_{C\cm}^\jet$ for the jet veto.
For all our cross section predictions we employ the MSTW 2008 PDFs~\cite{Martin:2009iq} together with their corresponding default value of $\alpha_s(m_Z)$.
We use LO PDFs at NLL and NLO PDFs at NLL$'+$NLO, such that the PDF order agrees with the perturbative order of the FO cross section components.
For all our results, we set $m_H = 125\GeV$ and $E_\cm = 8\TeV$, except for the comparison of the $\TauC^\jet$-binned cross section to the ATLAS data, where we use $m_H = 125.4\GeV$ as in the measurement.

We first display the resummed NLL$^\prime$ and nonsingular NLO contributions separately and the full NLL$'+$NLO results given by their sum for the jet-vetoed (cumulant) cross sections as a function of $\Tau^\cut$ in \fig{NLLpcentral}. In the left and middle panels we compare the two observables of $\Tau_{B,C}^\jet$ type, respectively. In the right panel of \fig{NLLpcentral} we compare the same results for $\TauB^\jet$ and $\TauC^\jet$ with each other.

The NLL$^\prime$ resummed contribution to the $\TauBcm^\jet$ cumulant cross section (blue dashed curve) is larger than the one to the $\TauB^\jet$ veto cross section (blue solid curve).
This is due to the additional $Y$-dependent terms present in the factorization formula for $\TauBcm^\jet$ in \eq{TBCcmResum} (which with the resummation switched on also depend on $\Tau^\cut$).
As a consequence, the nonsingular contribution to the $\TauBcm^\jet$ cumulant (green dashed curve) is slightly more negative than the one for $\TauB^\jet$ (green solid curve), so that the sum of resummed and nonsingular contributions for each observable reproduces the same total cross section for $\Tau^\cut \to \infty$. This can be seen directly from the combined NLL$^\prime+$NLO cross section in solid and dashed dark orange curves that approach the same constant value for large $\Tau^\cut$. For $\Tau^\cut \to 0$ on the other hand the Sudakov resummation forces the NLL$^\prime+$NLO cross section to vanish.
Note that for the cross sections integrated over the full $Y$ range shown in \fig{NLLpcentral} the difference between the resummed NLL$^\prime+$NLO predictions with a $\TauBcm^\jet$ and a $\TauB^\jet$ veto is hardly visible.
The same discussion holds for the comparison between $\TauCcm^\jet$ and $\TauC^\jet$ shown in the middle panel of \fig{NLLpcentral}.

In the right panel of \fig{NLLpcentral} we see that the NLL$^\prime+$NLO cross section for $\TauC^\jet < \Tau^\cut$ is larger and approaches the total cross section sooner than the one for $\TauB^\jet < \Tau^\cut$.
This difference in the shape of the cumulants arises due to the larger constant term in the NLO soft function $S_{gg}^C$ than in $S_{gg}^B$ [cf. \eqs{TauBsoftfct}{TauCsoftfct}], which enters as part of the resummed NLL$^\prime$ contribution to the cross section, \eq{TBCResum}. Since the total cross section at large $\Tau^\cut$ has to be the same for both observables, the larger singular contribution for $\TauC^\jet$ must eventually be compensated for by its nonsingular contribution when integrated over a large enough range of $\TauC^\jet < \Tau^\cut$, which is indeed more negative than for $\TauB^\jet < \Tau^\cut$.

\begin{figure*}[t!]
\begin{center}
\includegraphics[width=0.69\columnwidth]{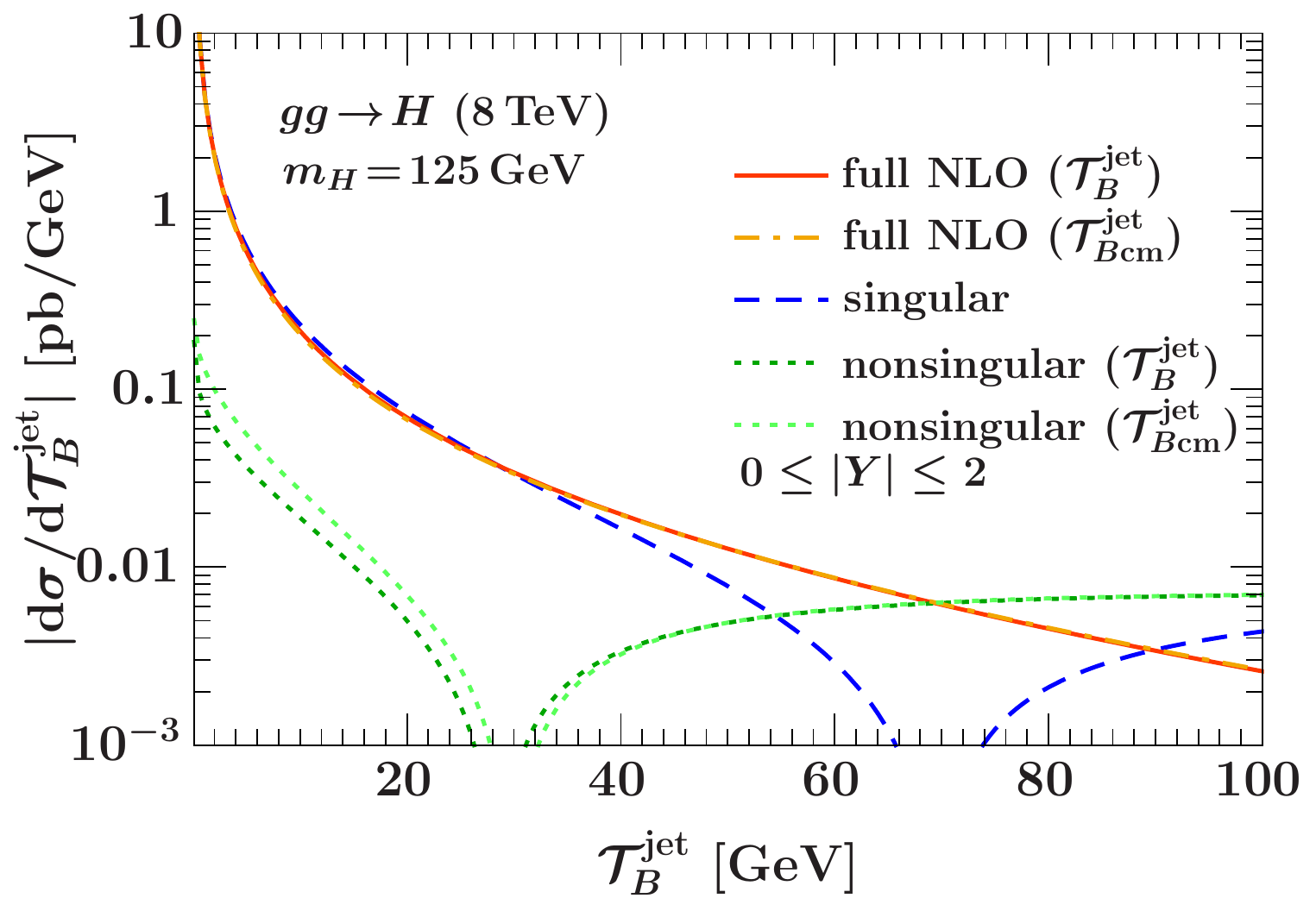}%
\hfill%
\includegraphics[width=0.69\columnwidth]{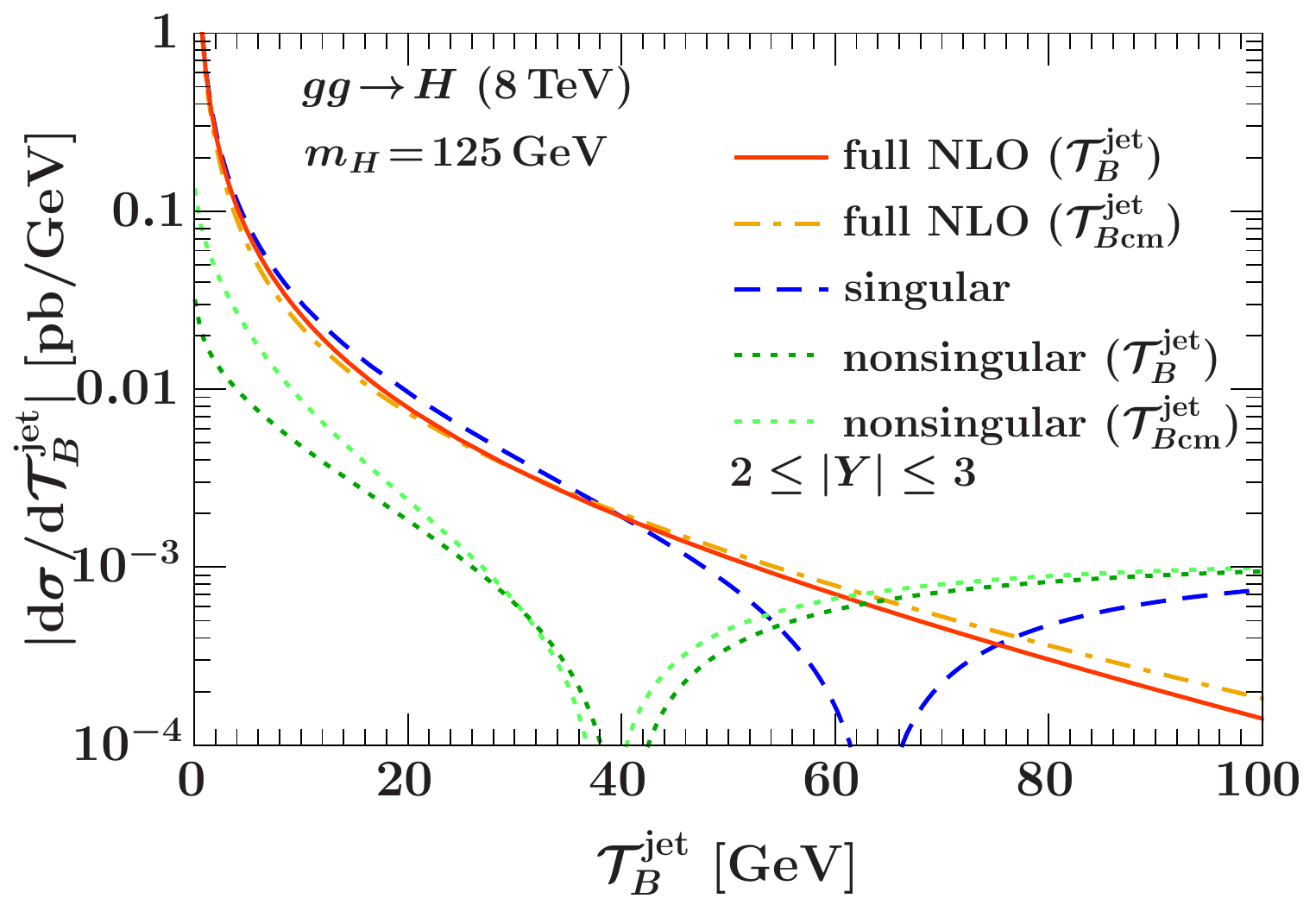}%
\hfill%
\includegraphics[width=0.69\columnwidth]{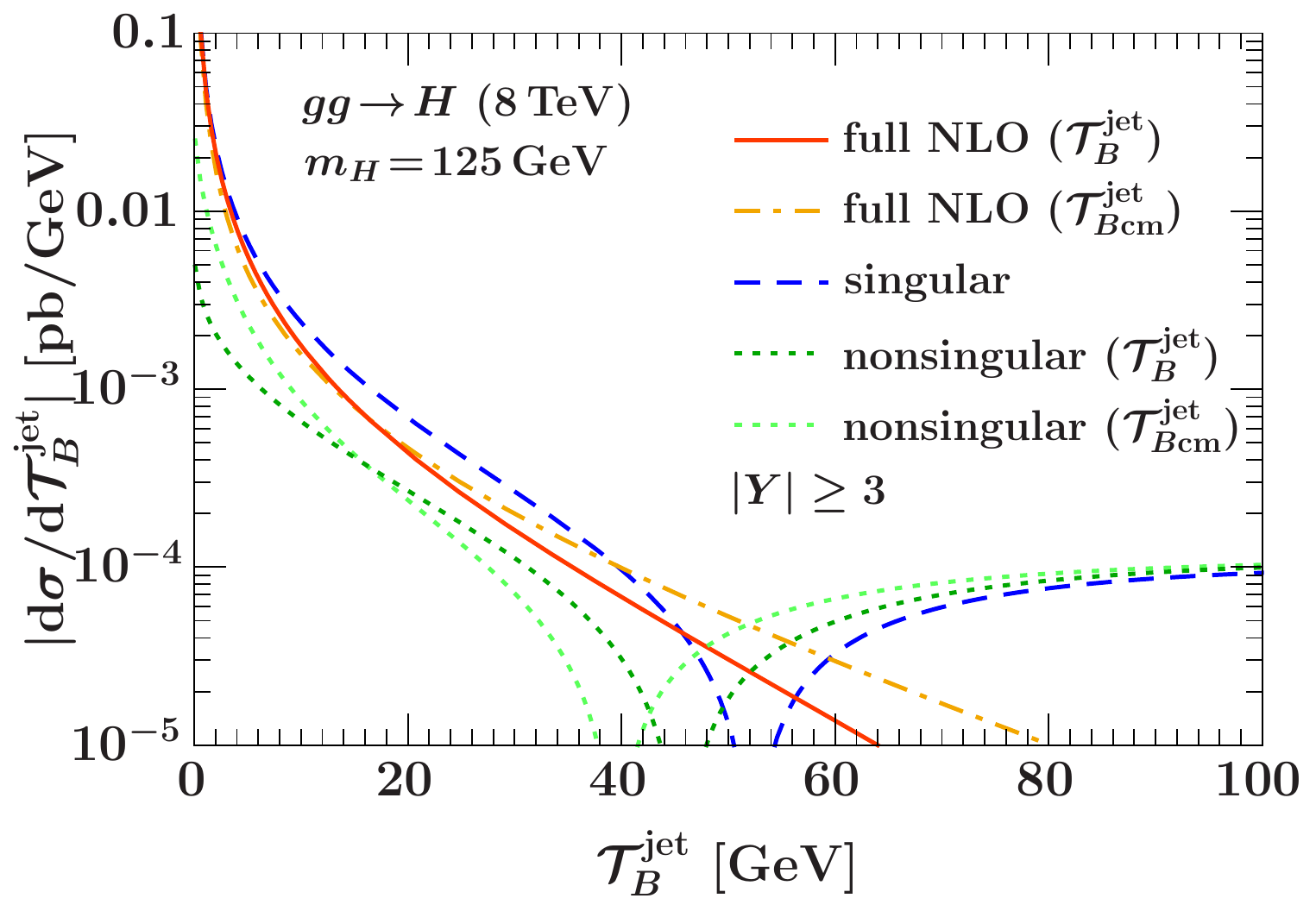}%
\end{center}
\vspace{-1ex}
\caption{Differential distributions for $\Tau_{B(\cm)}^\jet$ for the $\abs{Y}\le 2$ (left), $2\le \abs{Y}\le 3$ (middle), and $\abs{Y}\ge 3$ bins (right) to be compared with the left panel in \fig{Singnonsing}, where the cross sections have been integrated over the full $Y$ range.}
\label{fig:YdepSpectra}
\end{figure*}

\begin{figure*}[t!]
\begin{center}
\includegraphics[width=0.68\columnwidth]{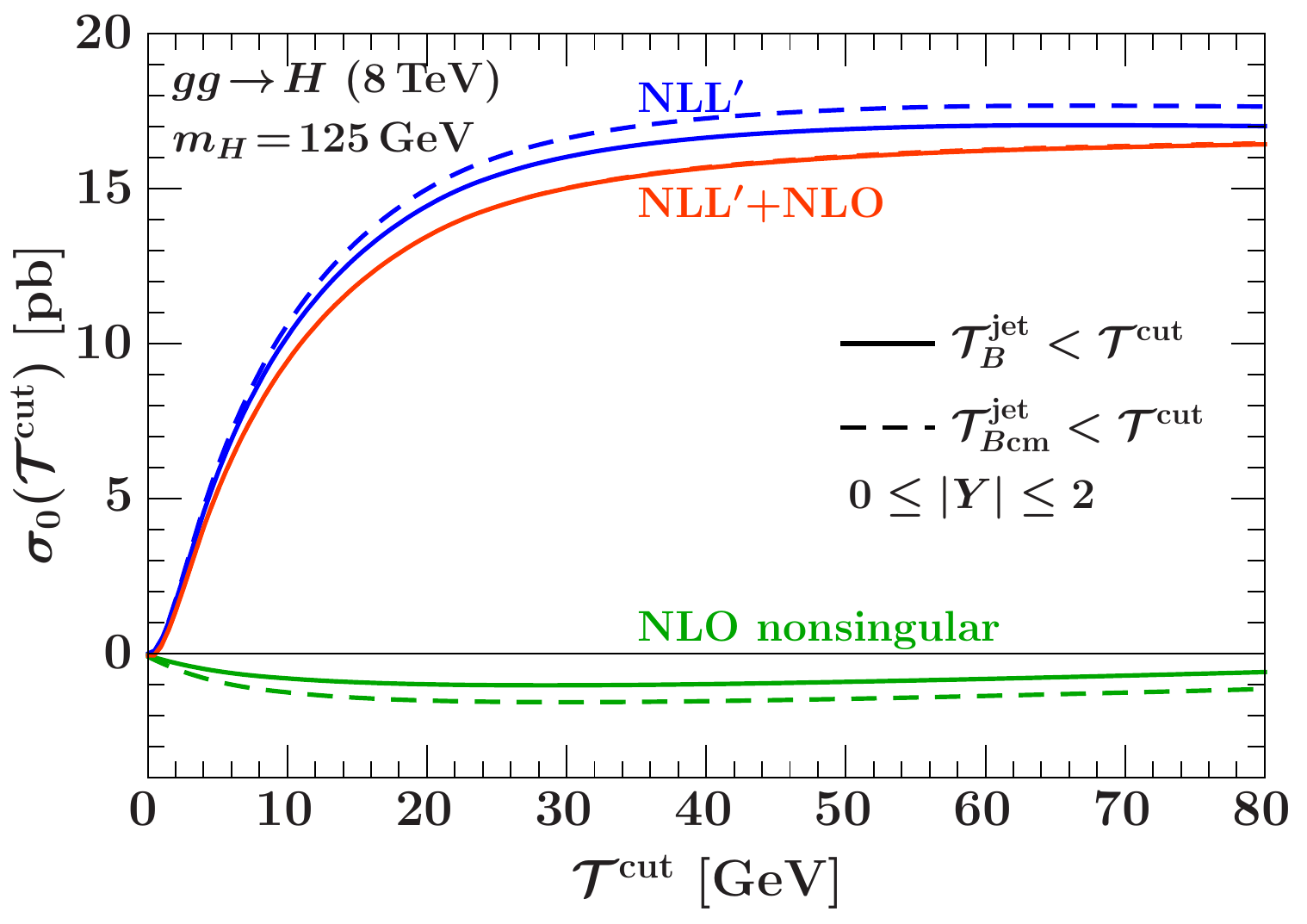}%
\hfill%
\includegraphics[width=0.68\columnwidth]{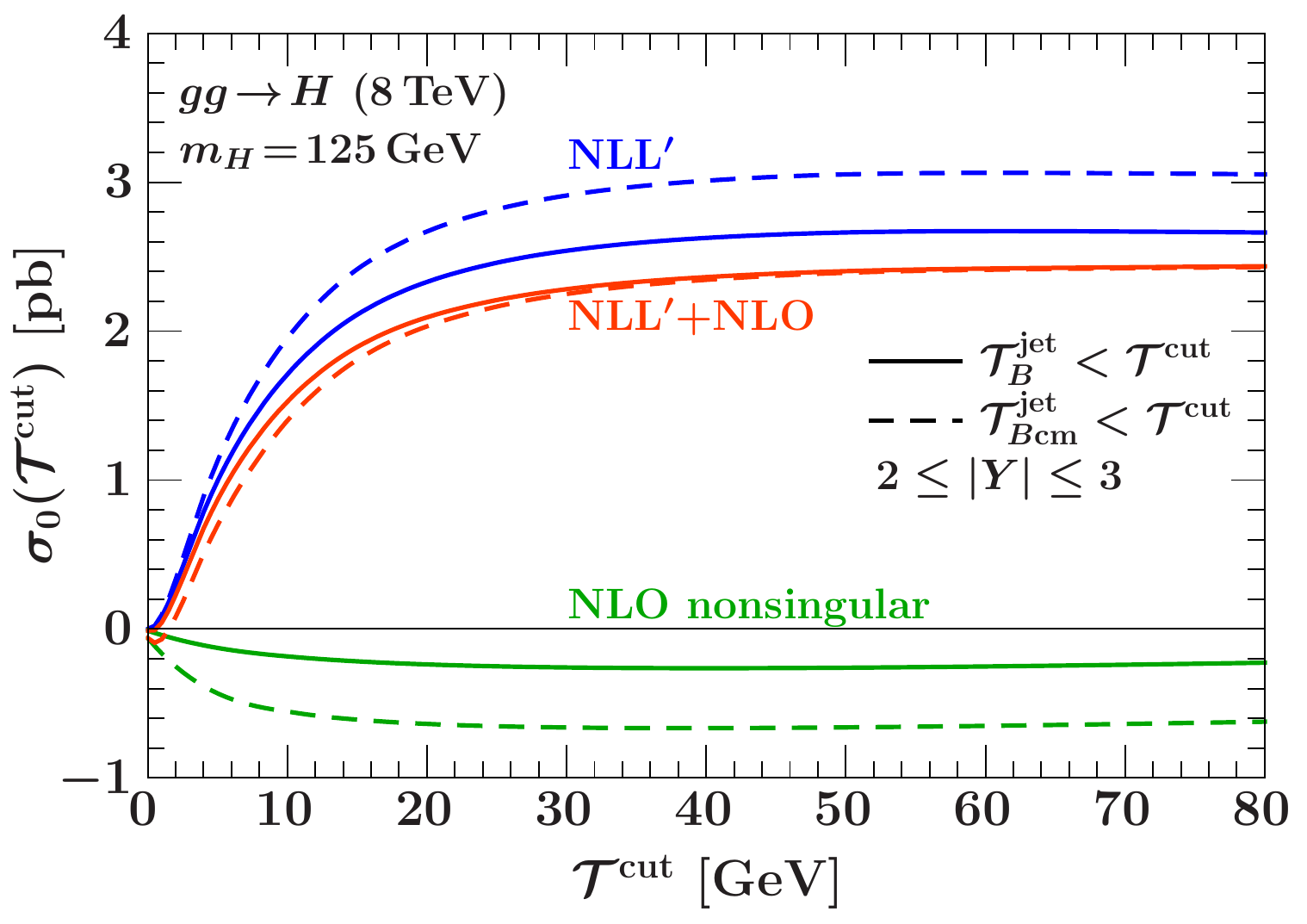}%
\hfill%
\includegraphics[width=0.71\columnwidth]{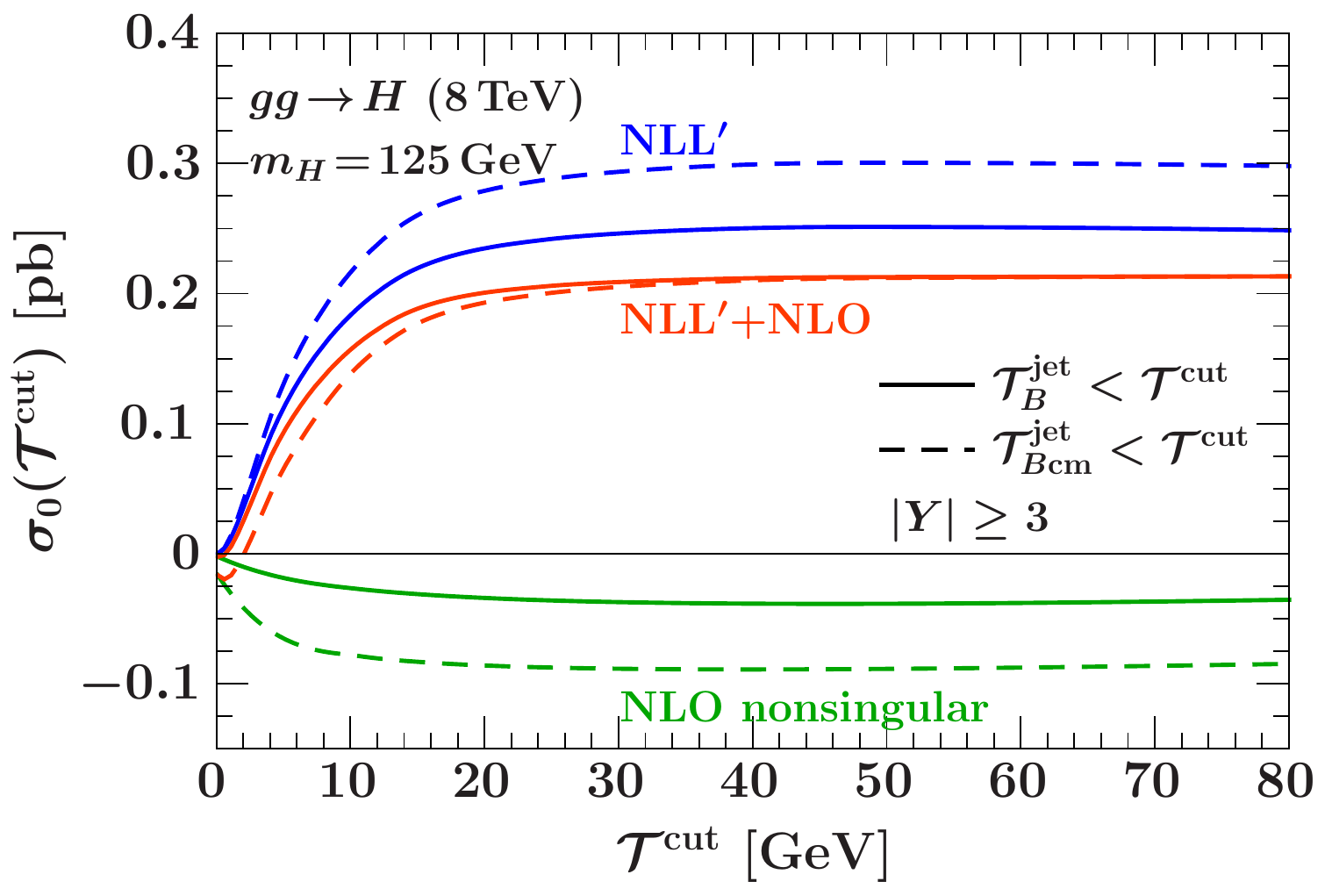}%
\end{center}
\vspace{-1ex}
\caption{Integrated $\Tau_{B(\cm)}^\jet$ cumulants for the $\abs{Y}\le 2$ (left), $2\le \abs{Y}\le 3$ (middle), and $\abs{Y}\ge 3$ bins (right) to be compared with the left panel in \fig{NLLpcentral}, where the cross sections have been integrated over the full $Y$ range.}
\label{fig:YdepCumulants}
\end{figure*}

To analyze the differences between the two versions of the $\Tau_{B,C}^\jet$-type variables in more detail, we compare in \figs{YdepSpectra}{YdepCumulants} the $\TauB^\jet$ and $\Tau_{B\cm}^\jet$ cross sections integrated over different $\abs{Y}$ ranges (bins).
The leftmost plots in \figs{YdepSpectra}{YdepCumulants} show the spectrum and cumulant cross sections for $\Tau_{B(\cm)}^\jet$ in the $\abs{Y}\le 2$ bin, respectively.
Qualitatively they look very similar to the corresponding plots for $\abs{Y}\le \ln (E_\cm/m_H)$ (i.e. the full $Y$ range),
except for the somewhat reduced total cross section in the left panel of \fig{YdepCumulants} due to the reduced $Y$ range.
In particular, the $\TauB^\jet<\Tau^\cut$ and $\TauBcm^\jet<\Tau^\cut$ vetoes again yield practically the same NLL$^\prime+$NLO cross sections.

The differences between $\TauB^\jet$ and $\TauBcm^\jet$ get more pronounced at larger Higgs rapidity $Y$. (The decrease in the overall normalization of all cross sections in \figs{YdepSpectra}{YdepCumulants} for larger $Y$ is due to the PDF suppression.) The $\Tau_{B(\cm)}^\jet$ spectra in \fig{YdepSpectra} show that the singular-nonsingular cancellations happen at lower $\Tau_B^\jet$ now, which means that following the discussion in \subsec{scalevar}, the parameters in the profile scales have to change accordingly.
For the resummed cross sections in the $2\le \abs{Y}\le 3$ and $\abs{Y}\ge 3$ bins shown in \fig{YdepCumulants}, we therefore set our profile parameters to $\{x_1, x_2, x_3\} = \{0.1,0.325,0.55\}$ and $\{x_1, x_2, x_3\} = \{0.1,0.275,0.45\}$, respectively.
As observed in the middle and right panels of \fig{YdepCumulants}, the $\TauB^\jet$ and $\TauBcm^\jet$ cumulants for the $2\le \abs{Y}\le 3$ bin start to differ in their shape: the separation between the respective resummed and nonsingular contributions is increased and the NLL$^\prime+$NLO $\TauBcm^\jet$ result considerably deviates from the one for $\TauB^\jet$ at small $\Tau^\cut$ values. These effects are even more enhanced for the $\abs{Y}\ge 3$ bin. 
For very small values of $\Tau^\cut$ ($\lesssim 2 \GeV$) the NLL$^\prime+$NLO $\TauBcm^\jet$ cumulants in the higher $Y$ bins turn slightly negative.
This unphysical effect is formally higher-order and due to a lack of Sudakov suppression of the large nonsingular corrections meaning that the resummation for $\TauBcm^\jet$ is less effective than for $\TauB^\jet$. 
The above conclusions from the analysis of the $Y$-binned cross sections likewise hold for the $\Tau_{C(\cm)}^\jet$ observables.

\begin{figure*}[t!]
\begin{center}
\includegraphics[width=\columnwidth]{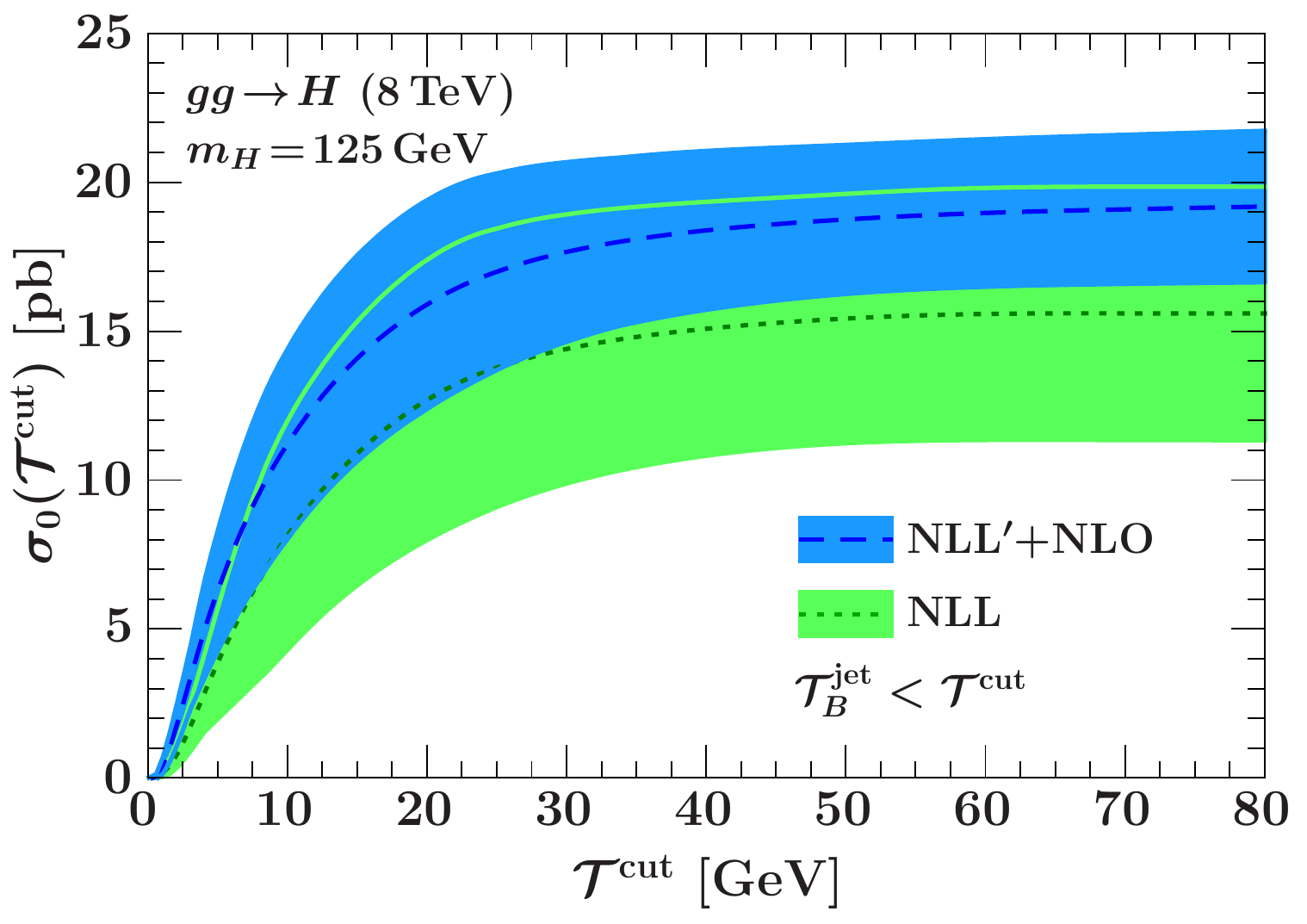}%
\hfill%
\includegraphics[width=\columnwidth]{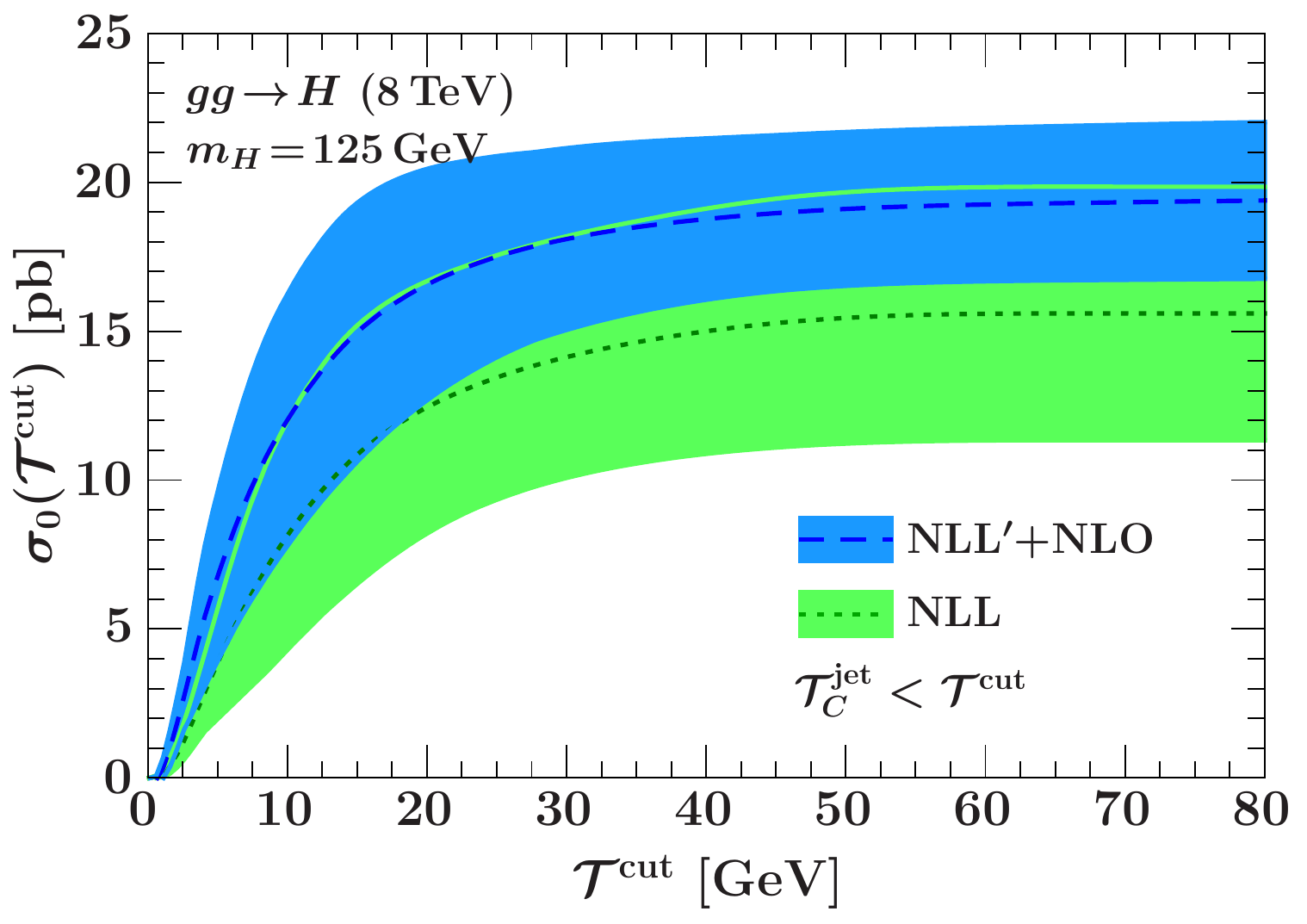}%
\end{center}
\vspace{-1ex}
\caption{Cumulant cross sections for $\TauB^\jet<\Tau^\cut$ (left panel) and $\TauC^\jet<\Tau^\cut$ (right panel). The (overlapping) green and blue bands represent the NLL and NLL$^\prime+$NLO predictions with the corresponding uncertainty ($\pm \Delta_0$) from scale variations according to \subsec{uncertainties}, respectively. The dashed and dotted lines indicate the respective default predictions using our central profiles for the resummation/FO scales as explained in \subsec{scalevar}.}
\label{fig:bandplots}
\end{figure*}

In \fig{bandplots}, we finally present the resummed Higgs+0-jet cross section predictions along with their perturbative uncertainty bands ($\pm \Delta_0$) obtained by the scale variations defined in \subsec{uncertainties}. 
To study the convergence of our resummed predictions and validate out uncertainty estimates, we show the NLL bands in green color and the  NLL$^\prime+$NLO bands in blue color for $\TauB^\jet < \Tau^\cut$ (left panel) and $\TauC^\jet < \Tau^\cut$ (right panel).

We observe a substantial decrease in uncertainties going from NLL to NLL$^\prime+$NLO, which is mostly due to NLO singular matching corrections, which partly cancel the scale variation from the NLL resummation factors.
Both NLL$^\prime+$NLO bands have an overlap with their NLL pendants that is consistent with our uncertainty estimates. 
We emphasize however that more solid conclusions about the order-by-order convergence of the perturbative predictions can be drawn once also the next higher order, i.e. NNLL$^\prime+$NNLO, is known, which is left for future work.

At $\Tau^\cut \sim 25 \GeV$, we find a perturbative uncertainty of about $20\%$ for our NLL$^\prime+$NLO predictions, which is largely driven by the sizeable FO uncertainties. It is also comparable with the precision obtained for the $p_T^\jet$-vetoed cross section at the same order in \mycite{Stewart:2013faa}.
Similar to the case of $p_T^\jet$, we also expect a substantial improvement in the precision for $\Tau_{B,C}^\jet$ when eventually going to NNLL$^\prime+$NNLO.
For $\Tau^\cut \sim 100\GeV$, we find a perturbative error of about $13\%$, which effectively equals the scale variation uncertainty of the total $\pi^2$-improved NLO cross section. 

\begin{figure}[t!]
\begin{center}
\includegraphics[width=\columnwidth]{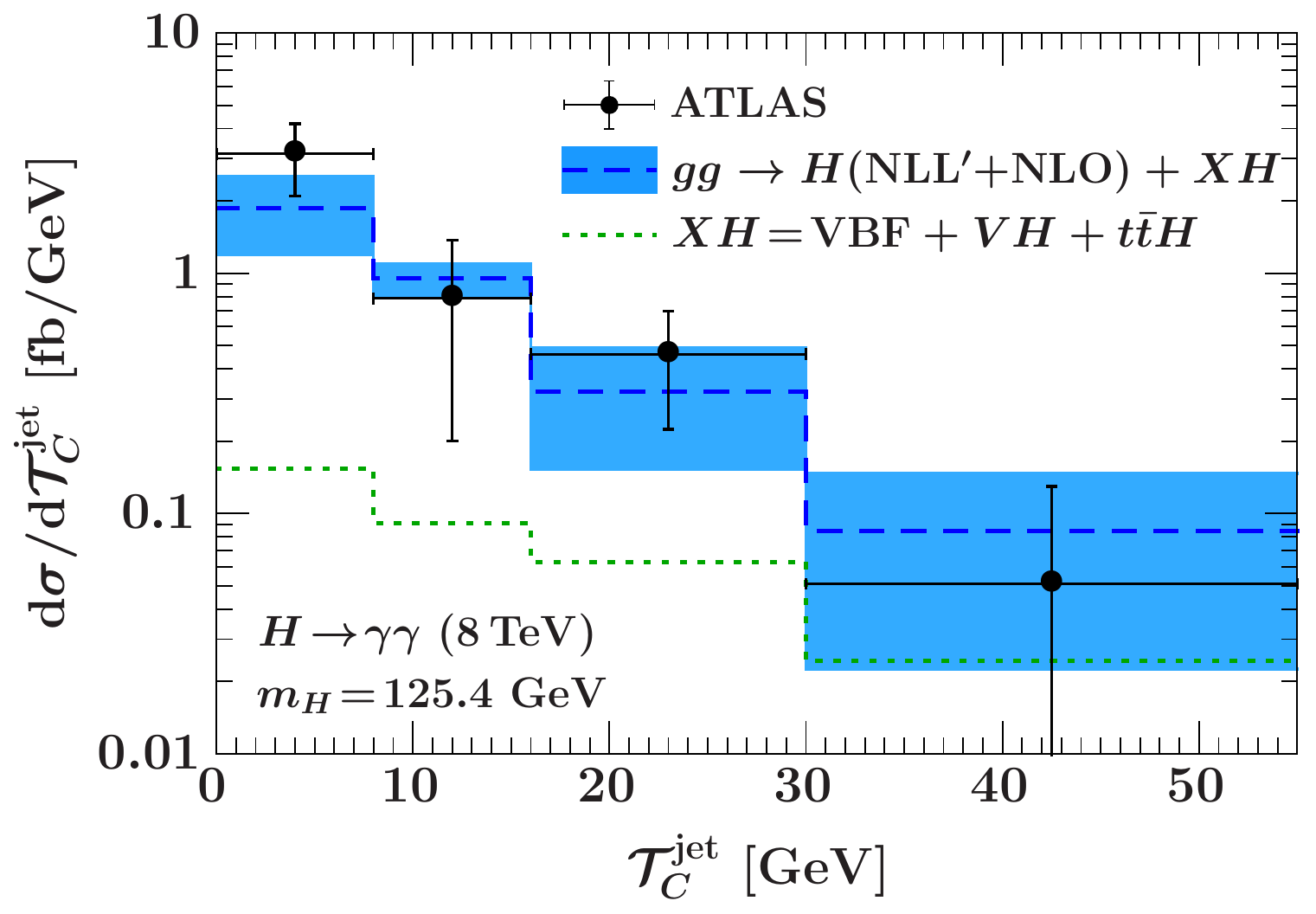}%
\hfill%
\end{center}
\vspace{-1ex}
\caption{Comparison of the $gg\to H \to \gamma\gamma$ cross section at NLL$'+$NLO in bins of $\TauC^\jet$ to the ATLAS $H\to\gamma\gamma$ measurements~\cite{Aad:2014lwa}. See the text for further details on the applied corrections.}
\label{fig:ATLAS}
\end{figure}

Last but not least, in \fig{ATLAS} we compare our prediction at NLL$^\prime+$NLO in bins of $\TauC^\jet$ with the recent ATLAS measurement in the $H \to \gamma \gamma$ channel~\cite{Aad:2014lwa}. Currently, the measurement has an underlying technical cut on reconstructed jets of $p_{Tj} \geq 25\GeV$, which effectively moves events between the first two bins. We correct for this effect by applying an extrapolation factor from Monte Carlo simulation.%
\footnote{We thank Marco Filipuzzi and Dag Gillberg for discussions on this point. At present this extrapolation introduces a nonnegligible MC model dependence. To minimize this in the future, it would be advantageous to try to reduce the lower cut on reconstructed jets as much as possible and/or slightly increase the lowest $\Tau_{Cj}$ bin edge.}
To directly compare with the measurements, we multiply our predictions by the $H\to\gamma\gamma$ branching ratio of $0.228\pm 0.011$~\cite{Heinemeyer:2013tqa} and apply several correction factors as given in Ref.~\cite{Aad:2014lwa}: The diphoton kinematic acceptance and photon isolation efficiency are essentially independent of $\TauC^\jet$, while nonperturbative corrections due to hadronization and underlying event are practically irrelevant.
Finally, we also add the contributions from other production channels as estimated in \mycite{Aad:2014lwa}, and which are shown by the green dotted lines. For the uncertainties in our NLL$'+$NLO predictions we propagate the $\Delta_\FO$ and $\Delta_\resum$ uncertainties by taking the differences of our cumulant predictions at the two bin edges separately for each profile variation. We also add in quadrature an $8\%$ uncertainty for PDF+$\alpha_s$ uncertainties (which we take to be the same as for the total cross section since they are mostly independent of $\Tau_{Cj}$).

\section{Conclusions}
\label{sec:conclusions}

In this paper, we have discussed the factorization and resummation of cross sections with rapidity-dependent jet vetoes.
Experimentally, such generalized jet vetoes have the advantage to provide efficient methods to veto central jets, while relaxing the phase space constraints (and therefore the requirements on the measurement) for jets with increasingly forward rapidities.
We introduced the jet-veto variables $\Tau_{B(\cm)}^\jet$ and $\Tau_{C(\cm)}^\jet$ which have two different types of rapidity weighting and related resummation properties. Since their resummation structure is notably different than for the $p_T^\jet$ observable, rapidity-weighted $\Tau_f^\jet$-binned cross sections yield valuable complementary information on the properties of additional jet production in a given hard process.

As a concrete example we considered Higgs$+$0-jet production through gluon fusion at the LHC, and presented cross section predictions at the NLL$^\prime+$NLO level for all four jet-veto variables. We analyzed their theoretical uncertainties via combined scale variations of the different involved resummation and FO scales. We find that the level of theoretical precision that can be reached for such rapidity-weighted jet-veto observables is comparable to what is currently possible for $p_T^\jet$ vetoes.
Comparing our analytic predictions for the $\TauC^\jet$-binned cross section with a recent ATLAS measurement in the $H\to\gamma\gamma$ channel~\cite{Aad:2014lwa} we find good agreement.

Hence, there are strong motivations that rapidity-dependent jet-vetoes, like the $\Tau_{B(\cm)}^\jet$ and $\Tau_{C(\cm)}^\jet$ variables discussed here, should be measured in other hadron collider processes such as Drell-Yan, diphoton, and weak diboson production at different invariant masses and rapidities of the produced color-singlet state. This will provide stringent tests of our understanding of jet-veto resummations and jet production in general. In turn, such generalized jet vetoes can be utilized to optimize signal selections in experimental analyses that rely on jet-binning, such as Higgs property measurements or new-physics searches.

\begin{acknowledgments}

We thank Florian Bernlochner, Marco Filipuzzi, Dag Gillberg, Philippe Gras, and Kerstin Tackmann for helpful conversations and/or comments on the manuscript.
This work was supported by the DFG Emmy-Noether Grant No. TA 867/1-1.

\end{acknowledgments}

\appendix

\section{Hard function}
\label{app:hard}

The hard function is defined as
\begin{align}
H_{gg}(m_t, q^2, \mu) = \abs{C_{ggH}(m_t, q^2, \mu)}^2
\,,\end{align}
where $C_{ggH}$ is the Wilson coefficient from matching the full $ggH$ form factor in the SM onto the $ggH$ current in SCET. For on-shell Higgs production it is evaluated at $q^2=m_H^2$. At one loop,
\begin{align} \label{eq:CggH}
&C_{ggH}(m_t, q^2, \mu)
\nn \\ & \quad
= \alpha_s(\mu)F^{(0)}\Bigl(\frac{q^2}{4m_t^2}\Bigr) \biggl\{
1 +\frac{\alpha_s(\mu)}{4\pi}
\biggl[C^{(1)}\Bigl(\frac{-q^2-\img 0}{\mu^2}\Bigr)
\nn\\ &\qquad
+ F^{(1)}\Bigl(\frac{q^2}{4m_t^2}\Bigr)\biggr]
\biggr\}
,\end{align}
where the coefficients $C^{(i)}$ and $F^{(i)}$ up to $i=2$ can be found in \mycite{Berger:2010xi}.
For our NLL$^\prime$ resummed predictions we need the NLO coefficient
\begin{align}
C^{(1)}(x) = C_A \Big(\!-\ln^2{x} + \frac{\pi^2}{6} \Big)\,,
\end{align}
and we used
\begin{align}
 F^{(0)}(z)&=\frac3{2z}-\frac3{2z} \Big|1-\frac1z \Big| \arcsin^2(\sqrt{z})\,, \\
 F^{(1)}(z)&=5C_A-3C_F + \ord{z}\,, \label{eq:F1}
\end{align}
where the terms neglected in \eq{F1} have a numerically very small effect as $m_H^2 \ll 4m_t^2$.

\section{Beam function}
\label{app:beam}

We expand the beam function matching coefficients in \eq{BOPE} as
\begin{align}
\cI_{ij}(t^\cut, z, R,\mu) = \sum_{n=0}^{\infty} \Bigl[\frac{\alpha_s(\mu)}{4\pi}\Bigr]^n \cI_{ij}^{(n)}(t^\cut, z, R,\mu)
\,.\end{align}
At tree level we have
\begin{align}
\cI_{ij}^{(0)}(t^\cut\!,z,R,\mu) = \delta_{ij}\, \delta(1-z)\,.
\end{align}
As explained in \subsec{resum}, we can obtain the one-loop matching coefficients for the gluon beam function by integrating the beam thrust matching coefficients $\cI_{gj}(t,z,\mu)$ in \mycite{Berger:2010xi} over $t$ from $0$ to $t^\cut$.
The results read
\begin{align}
\cI_{gg}^{(1)}(t^\cut\!,z,R,\mu) &= 2 C_A \theta(z) \Big[ \ln^2\frac{t^\cut}{\mu^2} \,\delta(1-z) \nn\\ 
&\quad + P_{gg}(z)\ln \frac{t^\cut}{\mu^2} + I_{gg}(z) \Big]
\label{eq:Igg}
\end{align}
and
\begin{align}
\cI_{gq}^{(1)}(t^\cut\!,z,R,\mu) &= 2 C_F \theta(z) \Big[ P_{gq}(z)\ln{\frac{t^\cut}{\mu^2}} + I_{gq}(z) \Big],
\label{eq:Igq}
\end{align}
where
\begin{align} \label{eq:Igdel_results}
I_{gg}(z)
&= \cL_1(1-z)\,\frac{2(1-z + z^2)^2}{z} - \frac{\pi^2}{6} \delta(1-z) \nn \\
&\quad  - P_{gg}(z) \ln z
\,,  \\
I_{gq}(z)
&= P_{gq}(z)\ln \frac{1-z}{z} + \theta(1-z) z 
\,.\end{align}
The LO gluon splitting functions are defined as
\begin{align} \label{eq:Pij}
P_{gg}(z)
&= 2 \cL_0(1-z) \frac{(1 - z + z^2)^2}{z}
\,,\nn\\
P_{gq}(z) &= \theta(1-z)\, \frac{1+(1-z)^2}{z}
\,,\end{align}
and
\begin{align} \label{eq:plusdefmaintxt}
\cL_n(x)
&= \biggl[ \frac{\theta(x) \ln^n x}{x}\biggr]_+
 = \lim_{\eps \to 0} \frac{\df}{\df x}\biggl[ \theta(x- \eps)\frac{\ln^{n+1} x}{n+1} \biggr]
\end{align}
denotes the usual plus distributions.

\section{Soft functions}
\label{app:soft}

\subsection{Soft function for $\Tau_{B(\cm)}^\jet$}
\label{app:softTauB}

The soft function for $\Tau_{B(\cm)}^{\jet}$ can be obtained by integrating the beam thrust soft function $S_{gg}(k,\mu)$ in \mycite{Berger:2010xi} over $0<k<\Tau^{\cut}$. Through NLO this yields
\begin{align} \label{eq:TauBsoftfct}
S_{gg}^B(\Tau^{\cut}\!,\mu) = 1 + \frac{\alpha_s(\mu)C_A}{\pi} \Bigl( -2 \ln^2{\frac{\Tau^{\cut}}{\mu}} + \frac{\pi^2}{12} \Bigr)
\,.\end{align}

\subsection{Soft function for $\Tau_{C(\cm)}^\jet$}
\label{app:softTauC}
In this appendix we calculate the soft function $S_{gg}^C$ at one loop.
The bare one-loop soft function for a generic (differential) measurement function $\cM(\Tau)$ and two (incoming) gluons is given by~\cite{Jouttenus:2011wh}
\begin{align}\label{eq:Soft1loop}
S_{gg}^{\bare(1)}(\Tau) &= 4 C_A\, g^2 \Big(\frac{e^{\gamma_E}\mu^2}{4\pi}\Big)^{\!\eps} \!\int \!\!\frac{d^dp}{(2\pi)^d} (p^+p^-)^{-1} \nn\\
& \quad \times 2\pi \delta(p^2)\theta(p_0) \, \cM(\Tau,p^+,p^-)\,,
\end{align}
where $p$ is the momentum of the emitted soft gluon.

At one loop the soft measurement function for $\Tau_C^\jet$ according to \eq{TauC} reads
\begin{align} \label{eq:MTauC1loop}
\cM (\Tau_C^\jet, p^+, p^-)
&= \delta \Bigl(\Tau_C^\jet - \frac{|\vec{p}_T|}{e^Y + e^{-Y}}\Bigr)
\nn\\
&=\delta \Bigl(\Tau_C^\jet - \frac{p^+ p^-}{p^+ + p^-}\Bigr)
\,.\end{align}
Inserting this in \eq{Soft1loop} and simplifying we get 
\begin{align}
S^{\bare(1)}_{gg}(\Tau_C^\jet) &= \frac{\alpha_s C_A}{\pi} \frac{(e^{\gamma_E}\mu^2)^\epsilon}{\Gamma(1-\epsilon)}\int \!\! dp^+ dp^- \frac{\theta(p^+)\theta(p^-)}{(p^+ p^-)^{1+\epsilon}} \nn\\
& \quad \times \delta \Bigl(\Tau_C^\jet - \frac{p^+ p^-}{p^+ + p^-}\Bigr)
\,.\end{align}
Integration over $p^+$ and $p^-$ yields
\begin{align}
S^{\bare(1)}_{gg}(\Tau_C^\jet)&= -\frac{\alpha_sC_A}{\pi}\frac{(e^{\gamma_E}\mu^2)^\eps}{\Gamma(1-\eps)} \frac{\Gamma(\eps)^2}{\Gamma(2\eps)} (\Tau_C^\jet)^{-1-2\eps} \,. 
\end{align}
Expanding $(\Tau_C^\jet)^{-1-2\eps}$ in terms of plus distributions and subtracting the $1/\epsilon$ divergence, the $\msb$ renormalized one-loop piece of the differential soft function reads
\begin{align}
S^{(1)}_{gg}(\Tau_C^\jet)
&= \frac{\alpha_s C_A}{\pi} \biggl[- \frac{4}{\mu}\mathcal{L}_1 \Big(\frac{\Tau_C^\jet}{\mu}\Big) + \frac{\pi^2}{4}\delta(\Tau_C^\jet) \biggr]
\,. \end{align}
As expected, replacing $C_A\to C_F$, this result agrees with the one-loop soft function for the C-parameter event shape in $e^+e^-\to q\bar q$~\cite{Alioli:2012fc, Hoang:2014wka}.
For the $\Tau_C^\jet$-veto we integrate over $\Tau_C^\jet$ and find
\begin{align} \label{eq:TauCsoftfct}
S_{gg}^C(\Tau^{\cut}\!,\mu) = 1 + \frac{\alpha_s(\mu)C_A}{\pi} \Bigl(-2 \ln^2{\frac{\Tau^{\cut}}{\mu}} + \frac{\pi^2}{4} \Bigr)
\,,\end{align}
where we also added the trivial tree-level contribution.

This result only differs from the one in \eq{TauBsoftfct} for $\Tau_B^\jet$ in the $\Tau^\cut$-independent constant, while the logarithmic term is dictated by the RG structure and according to the discussion in \subsec{resum} is the same for all four observables we consider.

\section{Renormalization group evolution}
\label{app:RG}

Analogous to the RG-evolved beam function in \eq{Bresum}, we can write the RG-evolved hard and soft functions as
\begin{align}
H_{gg}(m_t,q^2,\mu) &= U_H(q^2,\mu_H, \mu)\, H_{gg}(m_t,q^2,\mu_H)
\,, \\
S_{gg}^B(\Tau^{\cut}\!,\mu) &= U_S(\Tau^{\cut},\mu_S,\mu)\, S_{gg}^B(\Tau^\cut,\mu_S)
\,,\end{align}
with the corresponding RG evolution factors given by
\begin{align}
U_S(\Tau^{\cut}\!,\mu_S,\mu) &= e^{K_S(\mu_S,\mu)}\Bigl(\frac{\Tau^{\cut}}{\mu_S}\Bigr)^{\eta_S(\mu_S,\mu)}\,, \label{eq:US} \\
U_H(q^2,\mu_H, \mu) &= \Big| e^{K_H(\mu_H,\mu)}\Big(\frac{-q^2\!-\!\img0}{\mu_H^2}\Big)^{\!\!\eta_H(\mu_H,\mu)} \Big|^2 \!. \label{eq:UH}
\end{align}
The functions $K_i$ and $\eta_i$ in Eqs.~\eqref{eq:evolveB},~\eqref{eq:US} and~\eqref{eq:UH} are defined as 
\begin{align} 
K_B(\mu_B,\mu) &= +4 K_{\Gamma}^g(\mu_B,\mu) + K_{\gamma_B^g}(\mu_B,\mu)
\,,  \nn \\
K_S(\mu_S,\mu) &= -4 K_{\Gamma}^g(\mu_S,\mu)  + K_{\gamma_s^g}(\mu_S,\mu)
\,,  \nn \\
K_H(\mu_H, \mu) &= -2 K_\Gamma^g(\mu_H,\mu) + K_{\Gamma_H^g}(\mu_H,\mu)
\,,\end{align}
and
\begin{align}
\eta_B(\mu_B,\mu) &= -2\eta_{\Gamma}^g(\mu_B,\mu)\,, \nn\\
\eta_S(\mu_S,\mu) &= +4\eta_{\Gamma}^g(\mu_S,\mu)\,, \nn\\
\eta_H(\mu_H,\mu) &= +\eta_\Gamma^g(\mu_H,\mu)\,, 
\end{align}
where
\begin{align}
K_{\Gamma}^g(\mu_0,\mu)&=\int_{\alpha_s(\mu_B)}^{\alpha_s(\mu)} \frac{d\alpha}{\beta[\alpha]}\Gamma_{\cusp}^g[\alpha] \int_{\alpha_s(\mu_B)}^{\alpha} \frac{d\alpha}{\beta[\alpha]} , \nonumber\\
K_{\gamma_i^g}(\mu_0,\mu) &= \int_{\alpha_s(\mu_B)}^{\alpha_s(\mu)} \frac{d\alpha}{\beta[\alpha]} \gamma_i^g[\alpha], \nonumber\\
\eta_{\Gamma}^g(\mu_0,\mu) &=\int_{\alpha_s(\mu_B)}^{\alpha_s(\mu)} \frac{d\alpha}{\beta[\alpha]} \Gamma_{\cusp}^g[\alpha]\,. 
\end{align}
Explicit analytic expressions for the $K_\Gamma^g$, $K_{\gamma_i^g}$ and $\eta_i^g$ as well as the anomalous dimensions and beta functions relevant for NLL resummation can be found e.g. in Appendix B.3 of \mycite{Berger:2010xi}.


\bibliographystyle{../physrev4}
\bibliography{../beamfunc}

\end{document}